\documentclass[structabstract]{aa}  
\usepackage[]{natbib}
\usepackage{graphicx}
\usepackage{txfonts}
\usepackage{multirow}
\usepackage{graphicx}
\usepackage{epsfig}
\usepackage{txfonts}
\usepackage{psfrag}
\usepackage{units}
\DeclareMathAlphabet{\mathpzc}{OT1}{pzc}{m}{it}
\def\sun{\hbox{$\odot$}}
\def\nh{{n_{\rm H}}}

\def\nh2{{n(\rm H_2)}}
\def\h2{${\rm H_2}$}

\def\3cm{\rm {cm^{-3}}}
\def\2cm{\rm {cm^{-2}}}
\def\s-1{\rm {s^{-1}}}
\def\etal {et al.}

\def\Msun{$M_{\sun}$}
\def\mum {\hbox{$\mu\rm m$}}
\def\kms {\hbox{${\rm km\,s}^{-1}$}}

\def\ergs{{\rm {erg}~{s^{-1}}}}

\def\Wcmusr{{\rm {W}~{{cm^{-2}}~{{\mu m}^{-1}}~sr^{-1}}}}
\def\Wcmsr{{\rm {W}~{{cm^{-2}}~sr^{-1}}}}
\def\Wcm{{\rm {W}~{{cm^{-2}}}}}
\def\Wsr{{\rm {W}~sr^{-1}}}

\def\hcn{{{HCN}}}

\def\h2s1{{{H$_2$}S(1)}}
\def\h2s2{{{H$_2$}S(2)}}
\def\twco{{$^{12}$CO}}
\def\thco{{$^{13}$CO}}
\def\ceo{{C$^{18}$O}}

\def\oiii{{[O {\small III}]}}
\def\oiv{{[O {\small IV}]}}

\def\cli{{[Cl {\small I}]}}
\def\clii{{[Cl {\small II}]}}

\def\neii{{[Ne~{\small II}]}}
\def\neiii{{[Ne~{\small III}]}}

\def\nev{{[Ne~{\small V}]}}

\def\siii{{[S~{\small III}]}}
\def\siv{{[S~{\small IV}]}}

\def\SiIV{{[Si {\small IV}]}}
\def\SiIX{{[Si {\small IX}]}}

\def\hi{{H~{\small I}}}
\def\hii{{H~{\small II}}}
\def\c18o{{C$^{18}$O}}

\begin{document}
   \title{The deeply obscured AGN of NGC~4945\thanks{We dedicate this paper to the memory of our esteemed 
colleague and advisor, Alan Moorwood (1945-2011), who pioneered the infrared spectroscopic study of
NGC~4945. This work is based on observations obtained with the Spitzer Space Telescope, which is operated by the Jet Propulsion Laboratory, California Institute of Technology, under NASA contract 1407.}}

   \subtitle{I. Spitzer-IRS maps of \nev, \neii, H$_2$ 0-0 S(1), S(2), and other tracers}

   \author{J.P.~P\'erez-Beaupuits
          \inst{1}\fnmsep\inst{3}
          \and
          H.W.W.~Spoon\inst{2}
          \and
          M.~Spaans\inst{3}
          \and
          J.D.~Smith\inst{4}
          }

   \institute{Max-Planck-Institut f\"ur Radioastronomie, Auf dem H\"ugel 69, 53121 Bonn, Germany; 
              \email{jp@mpifr.de}
         \and
             Astronomy Department, Cornell University, Ithaca, NY
         \and
             Kapteyn Astronomical Institute, Rijksuniversiteit Groningen, 9747 AD Groningen, The Netherlands
         \and
             Department of Physics and Astronomy, Mail Drop 111, University of Toledo, 2801 West Bancroft Street, Toledo, OH 43606, USA 
             }

   \date{Received January 15, 2011; accepted March 15, 2011}

 
  \abstract
   {The nearly edge-on galaxy NGC~4945 is one of the closest galaxies where an AGN and starburst coexist, and is one of the brightest sources at $100~\rm keV$. Near and mid-infrared spectroscopy have shown very strong obscuration of its central region, rivaled only in strength by some of the most deeply obscured ULIRGs. 
   }
   {
   Determine the spatial distribution of ISM emission features in the central 426$\times$426 $\rm pc^2$ of NGC~4945.
   }
   {We map the central region of NGC~4945 in three of the four Spitzer-IRS modules (SH, SL and LL). In particular, we produce maps of the flux distribution of the starburst tracers \neii, \neiii, \siii\ and \siv\ at 12.81, 15.56, 18.71 and 10.51 \mum, respectively, and a map of the AGN narrow-line region tracer \nev\ at 14.32 \mum.
In addition, we mapped the S(1), S(2) and S(3) pure rotational lines of H$_2$, which trace the distribution of warm molecular hydrogen. 
Finally, we obtained an extinction map ($A_{\rm V}$) based on the apparent strength of the 9.7~\mum\ silicate absorption feature.}
   {
At a spatial resolution of $\sim$5$''$ our extinction map traces the contours of the starburst ring. The highest extinction is, however, found at the nucleus, where we measure $A_{\rm V}(9.85~\mu \rm m)$$\approx$60. Within the uncertainty of the astrometry all emission lines are found to peak on the nucleus, except for the warm molecular hydrogen emission which shows a maximum 60-100 pc NW of the nucleus. We favour a scenario in which the lower H$_2$ 0-0 S(1) and S(2) rotational lines originate mainly from an unobscured extra-nuclear component associated with the super-wind cone observed in the HST NICMOS map of the H$_2$ 1-0 S(1) vibrational line. 
For the \nev\ emission we infer an attenuation of a factor 12-160 ($A_{\rm V}$=55-112) based on a comparison of the ratio of our \nev\ flux and the absorption-corrected 14--195 keV Swift-BAT flux to the average \nev/BAT ratio for Seyfert 1 nuclei. The high attenuation indicates that \nev\ and \oiv\ cannot be used as extinction-free tracers of AGN power in galaxies with deeply buried nuclei.}
   {}


   \maketitle
%

\section{Introduction}\label{sec:intro}

At a distance of $\sim$3.82 Mpc ($\sim$18.5 pc/arcsec, \citealt{karachentsev07}) the active galaxy NGC~4945 is one of the closest galaxies that host both, an AGN and starburst. Earlier X-ray observations showed evidence for a hidden AGN \citep{iwasawa93, guainazzi00}. These observations revealed a Compton-thick spectrum with an absorbing column density of $N_{\rm H}=2.4\times 10^{24}~\2cm$ \citep{guainazzi00}. The nucleus of NGC~4945 is one of the brightest extragalactic sources at $100~\rm keV$ \citep{done96}, and the brightest Seyfert 2 AGN at $>20~\rm keV$ \citep{itoh08}.

Rather than a point source marking the presence of the AGN in optical and near-infrared images, the nearly edge-on ($i\sim 80^o$) line of sight to the central region reveals evidence for strong and patchy extinction, which is especially apparent in HST-NICMOS $H-K$ maps \citep{marconi00}. A dust lane aligned along the major axis of the galactic disk obscures parts of the central region just southeast of the $K$-band peak. The $K$-band peak itself lies $\sim1''$ west (but, within the uncertainty) of the position of the H$_2$O mega maser \citep{greenhill97}, which we adopt as the location of the AGN. The mega maser emission allows to estimate the mass of the central black hole, $M_{\rm BH}\approx1.4\times10^6~\rm M_{\sun}$, which together with its inferred 0.1--200 keV bolometric luminosity  of $\sim$$2\times10^{43}~\ergs$ ($\sim5\times10^9~L_{\sun}$, \citet{guainazzi00}) indicates that the central engine radiates at $\sim$10\% of its Edington luminosity \citep{greenhill97, madejski00}.

Estimates from IRAS observations indicate that about 75\% of the total infrared luminosity of the galaxy (L$_{\rm IR}=2.4\times10^{10} ~L_{\sun}$) is generated within an elongated region of $12''\times9''$ (about $222\times167~\rm pc^2$) centered on the nucleus \citep{brock88}. The structure of this region, as shown in high resolution HST-NICMOS observations of the Pa$\alpha$ line, is consistent with a nearly edge-on starburst disk with a $4.5''$ ($\sim83~\rm pc$) radius \citep{marconi00}, embedded in a $r=8''$ ($\sim$150 pc) inclined (62 degrees) molecular disk \citep{chou07}.

Although the star formation and supernova rates in the nuclear region were originally estimated to be moderate ($\sim0.4$ \Msun\ $\rm yr^{-1}$ and $\sim0.05~\rm yr^{-1}$, respectively; \citealt{moorwood94}), more recent estimates based on high resolution (angular scale of $0.3~\rm pc$) radio observations and estimates of supernova remnant source counts, sizes and expansion rates, lead to a type II supernova rate of $>0.1(v/10^4)~\rm yr^{-1}$, and star formation rate (SFR) limits of $2.4(v/10^4)<{\rm SFR}(M\geq5~M_{\sun}) < 370~M_{\sun}~\rm yr^{-1}$, where $v$ is the shell radial expansion velocity in \kms\ \citep{lenc09}. These supernova and star formation rates are, within a factor of two, similar to those estimated in NGC~253 and M82 \citep{pedlar03, lenc06}, which are also nearly edge-on starburst galaxies, have similar distances close to $4~\rm Mpc$ and all show similar infrared luminosities \citep{rice88}. The impact of the central starburst on the circumnuclear region is large, as revealed by the presence of a conical cavity evacuated by a supernova driven wind \citep{moorwood96a, marconi00}. 
The HST-NICMOS images of the H$_2$ 1-0 S(1) line at $2.12$ \mum\ show that the edges of the cavity extend out to $5''$ ($\sim$93$\rm pc$) north from the NGC~4945 nucleus \citep{marconi00}.

Near and mid-infrared spectroscopy indicate that the ISM line of sight to the central region of NGC~4945 is very different than that of other nearby starburst galaxies, such as M~82 and NGC~253, which present similar inclination angles. 
The strong absorption features of both volatile (CO and CO$_2$) and refractory (H$_2$O) ices observed in the 2.4--5 \mum\ ISO-PHT-S spectrum indicate the presence of shielded cold molecular clouds obscuring the NGC~4945 nucleus \citep{spoon00, spoon03}. 
The detection of a strong absorption feature of XCN ice at 4.62~\mum\ by \citet{spoon00, spoon03} indicates that these molecular clouds have been processed in an energetic environment \citep{lacy84}, and processed ice is suggested to be a common characteristic of dense molecular material in star forming galactic nuclei \citep{spoon03}. A highly distorted PAH emission spectrum produced by a very deep 9.7 \mum\ silicate absorption feature is more evidence for the unusually strongly obscured nuclear region \citep{spoon00, brandl06}. This makes NGC~4945 a unique nearby laboratory to study an environment that can be found only in distant ULIRGs.


Previous mid-infrared spectroscopic observations could not confirm the presence of an AGN in NGC~4945. Just an upper limit of the 14.32 \mum\ \nev\ emission, considered a tracer of AGN narrow-line regions \citep[e.g.,][]{moorwood96b, genzel98}, was obtained from ISO-SWS observations \citep{spoon00}. The VLT-ISAAC observation of the 3.93 \mum\ \SiIX\ line, commonly observed in the soft X-ray photoionized gas of many Seyfert galaxies \citep{oliva94, lutz02}, resulted in a non-detection \citep{spoon03}. Only recent observations with the more sensitive IRS-SH spectrograph on Spitzer allowed the detection (although at a very faint level) of the 14.32 \mum\ \nev\ line towards the NGC~4945 nucleus \citep{bernards09}. The \nev14.3\mum/\neii12.8\mum\ flux ratio found in NGC~4945 is $\sim0.007$, which indicates an AGN contribution of less than a few percent using the diagram by \citet{farrah07} (their Fig. 16). This detection of \nev\ relative to \neii\ is about 10 times weaker than what was observed in other Seyfert galaxies like Mrk~266 and NGC~1365 \citep{bernards09}.

In this paper we study the mid-IR properties of the nuclear region of NGC~4945. Spitzer-IRS spectral mapping observations of a $23''\times23''$ (about 426$\times$426$~\rm pc^2$) region are presented. The mapping capabilities of Spitzer-IRS allow the study of a number of properties of the nuclear region in NGC~4945. For instance, the extent of the AGN coronal line region as traced by the \nev\ line, the extent of the crystalline silicate absorbing region (and a search for the sources responsible for the presence of crystalline silicates), disentangling PAH emission and silicate absorption along the line of sight, and the excitation and age of the circumnuclear-nuclear starburst based on line ratios of forbidden lines.
Although we actually used the Spitzer-IRS modules SH, SL and LL, in this work we present the most interesting results from the SH and SL spectral maps only, since we focus on the analysis of the spatial distribution of ISM emission and absorption features. We present for the first time the maps of the starburst tracers \neii\ 12.81 \mum, \neiii\ 15.56 \mum, \siii\ 18.71 \mum, and \siv\ 10.51 \mum, and a map of the AGN narrow-line region tracer \nev\ at 14.32 \mum, and analyze the ratios between some of these tracers as different diagnostics.
The organization of this article is as follows. In Sect.~\ref{sec:obs} we describe the observations and the data reduction. The maps obtained are presented in Sect.~\ref{sec:results}. The analysis of the data is presented in Sect.~\ref{sec:discuss}. The conclusions and final remarks are presented in Sect.~\ref{sec:final-remarks}.

\section{Observations and data reduction}\label{sec:obs}

We mapped the central region of NGC~4945 with the InfraRed Spectrograph\footnote{The IRS was a collaborative venture between Cornell University and Ball Aerospace Corporation funded by NASA through the Jet Propulsion Laboratory and the Ames Research Center.} (IRS - \citealt{houck04}) on board of the Spitzer Space Telescope \citep{werner04} through the guaranteed time observation (GTO) program P40018 (PI: H.W.W.~Spoon). The spectral maps were done in moderate resolution (R$\sim$600) with the Short-High (SH; 9.9 -- 19.6 \mum) IRS\footnote{http://ssc.spitzer.caltech.edu/irs/highleveloverview} module, and at low resolution (R$\sim$60--120) using the Short-Low (SL1; 7.4 -- 14.5 \mum, SL2; 5.2 -- 7.7 \mum) and Long-Low (LL1; 19.5 -- 38.0 \mum, LL2; 14.0 -- 21.3 \mum) IRS modules. 

The SH module was used to measure mainly the fine-structure emission lines (starburst tracers) \siv\ at 10.51 \mum, \neii\ at 12.81 \mum, \neiii\ at 15.56 \mum, \siii\ at 18.71 \mum, the AGN narrow-line region tracer \nev\ at 14.32 \mum, and the molecular hydrogen (pure rotational) lines, H$_2$ 0-0 S(2) and H$_2$ 0-0 S(1) at 12.3 \mum\ and 17.0 \mum, respectively. 
With the SL module we obtained maps of the H$_2$ S(3), and the silicate absorption feature at 9.7~\mum.


\subsection{SH map}

\begin{figure}[t]
  


  \hspace*{\fill}\includegraphics[width=8cm]{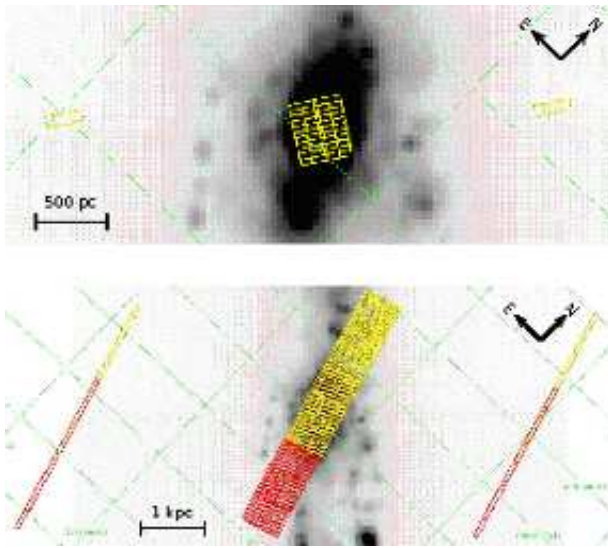}\hspace*{\fill}

  \caption{\footnotesize{Spitzer-MIPS 24 \mum\ map (Varoujian Gorjian, priv.comm.) of the central region of NGC~4945. The galaxy major axis runs vertical. North is toward the upper right, East toward the upper left. Along the declination axis the grid spacing is $\sim$1 arcminute, and about 12 seconds along the R.A. axis. The region mapped with the IRS-SH module (\textit{top panel}) is about $21''\times26''$, while a larger region of $37''\times67''$ was mapped with the IRS-SL (\textit{bottom panel}) module. The red and yellow grids correspond to the SL1 and SL2 spectral orders, respectively. The sky (off-source) positions observations are indicated by the grids to the left and right of the galaxy major axis. The images were produced with the Spitzer/Leopard software package used to query the archive and download the data.}}
  \label{fig:maps-overlay}
\end{figure}

The IRS-SH map was obtained in July 2007 (at the beginning of the summer visibility window) when the SH slit was relatively aligned with the minor axis of NGC~4945. The area mapped was limited to a set of 3 parallel by 10 perpendicular pointings centered on the nucleus, and covering an area of $21'' \times 26''$. 
The \textit{top panel} of Fig.~\ref{fig:maps-overlay} shows the orientation of the SH slit overlaid on the Spitzer-MIPS 24 \mum\ map of the central region of NGC~4945 (by courtesy of Varoujian Gorjian, private communication).

The quality of the map was optimized by splitting up the mapping into 4 1-cycle maps of 120 sec, each of which was alternated with a staring mode off-source sky measurement to keep track of rogue pixels and to obtain an accurate measure of the sky background. The total time used for the SH map, including sky positions, was 6.5 hrs.

\subsection{SL map}

Due to a \textit{Spitzer anomaly}\footnote{http://www.spitzer.caltech.edu/news/856-feature06-25-Engineers-Studying-Spitzer-Anomaly} that began a few hours before the original schedule of the observations, the IRS-SL map of the nuclear region of NGC~4945 was delayed till August 2008, when the SL slit was no longer perfectly aligned with the galaxy major axis.
The observations were done with 5 cycles of 14 sec ramps, which allowed to map a region of about $37''\times67''$ in 2 parallel by 19 perpendicular pointings centered on the nucleus of NGC~4945. 

In order to sample the background continuum and to identify rogue pixels, we also obtained off-source spectra with the same integration times and number of cycles at two positions sufficiently above and below the galaxy disk. The off-source spectra were taken before and after the SL mapping. The total time used for the SL map, including sky positions, amounts to 3.1 hrs.
The orientation of the SL slit, overlaid on the Spitzer-MIPS 24 \mum\ map of NGC~4945, is shown in the \textit{bottom panel} of Fig.~\ref{fig:maps-overlay}.




\subsection{Data reduction}\label{sec:data-reduction}

The Basic Calibrated Data (BCD) were pre-processed with the Spitzer pipeline versions S16.1 for SH, and S18.1 for SL. Rogue pixels were cleaned, and maps were built, using the IRS mapping reduction package CUBISM\footnote{http://ssc.spitzer.caltech.edu/dataanalysistools/tools/cubism/}, designed by the SINGS\footnote{http://ssc.spitzer.caltech.edu/spitzermission/observingprograms/legacy/} legacy team \citep{smith07}.
With this tool we performed the flux calibration, background subtraction, and estimated the statistical uncertainty at each spectral wavelength. We create the first spectral cubes from the slit observations combined with CUBISM. After creating the first cubes, we further cleaned the data using the back-tracking procedure described in \citet{smith07}. All the pixels with flux uncertainty larger than 50\% were flagged, so they were not used in subsequent reconstructions of the spectral cubes. We iterate on the cleaning procedure by performing visual inspections throughout the maps, of small sections of the spectra around the emission lines, PAH and silicate features of interest, and we reconstructed the cubes after cleaning new pixels.

The spectral cubes obtained with CUBISM are in units of surface brightness, $\rm MJy~sr^{-1}$, but we convert them to $\Wcmusr$ to work and present the spectra. We use the units of surface brightness $\Wcmsr$ to present the maps, and flux density $\Wcm$ for the integrated total fluxes. For this we use the conversion $1~\rm arcsec^2=2.3504\times10^{-11}~steradian$, knowing that the pixel size\footnote{See http://coolwiki.ipac.caltech.edu/index.php/Units for a description of Spitzer units and conversions} of the IRS/SH map is $\sim$2.26$~\rm arcsecs/pixel$, which leads to $1.2005\times10^{-10}~\rm steradians$ that we need to multiply by in order to get the maps in units of flux density. In the case of IRS/SL the pixel size is $\sim$1.85$~\rm arcsecs/pixel$, so we have slightly smaller $8.04424\times10^{-11}~\rm steradians$ per pixel than in the SH map.

\section{Analysis and results}\label{sec:results}

During the data reduction and spectral analysis process we found that the SH spectral orders\footnote{http://ssc.spitzer.caltech.edu/irs/irsinstrumenthandbook/} are mismatched at levels that vary across the mapped region. This effect is smoothed (although not completely solved) by increasing the overlap between the orders (in the wavsamp calibration file) and by increasing the size of the aperture used to extract 1-D average spectra, as discussed below. 
We also observed module to module mismatches, which inhibit us from reliably combining the SH and SL modules to perform further analyses like, for instance, using the H$_2$ S(3) line as an extinction indicator. These mismatches in the SH orders and between modules have not been addressed before\footnote{http://ssc.spitzer.caltech.edu/irs/features/}, probably because they are apparent only in high S/N data as those presented in this work. The IRS-SH/SL orders/modules mismatch problem seen in high S/N data is a new issue that we are still studying together with members of the SINGS team. Fortunately, most of the emission and absorption features we are interested in are found within the spectral orders, and those features observed near the edge of the orders are treated with extreme care or not addressed at all. 

\begin{figure}[htp]


  \hspace*{\fill}\includegraphics[width=8cm,angle=0]{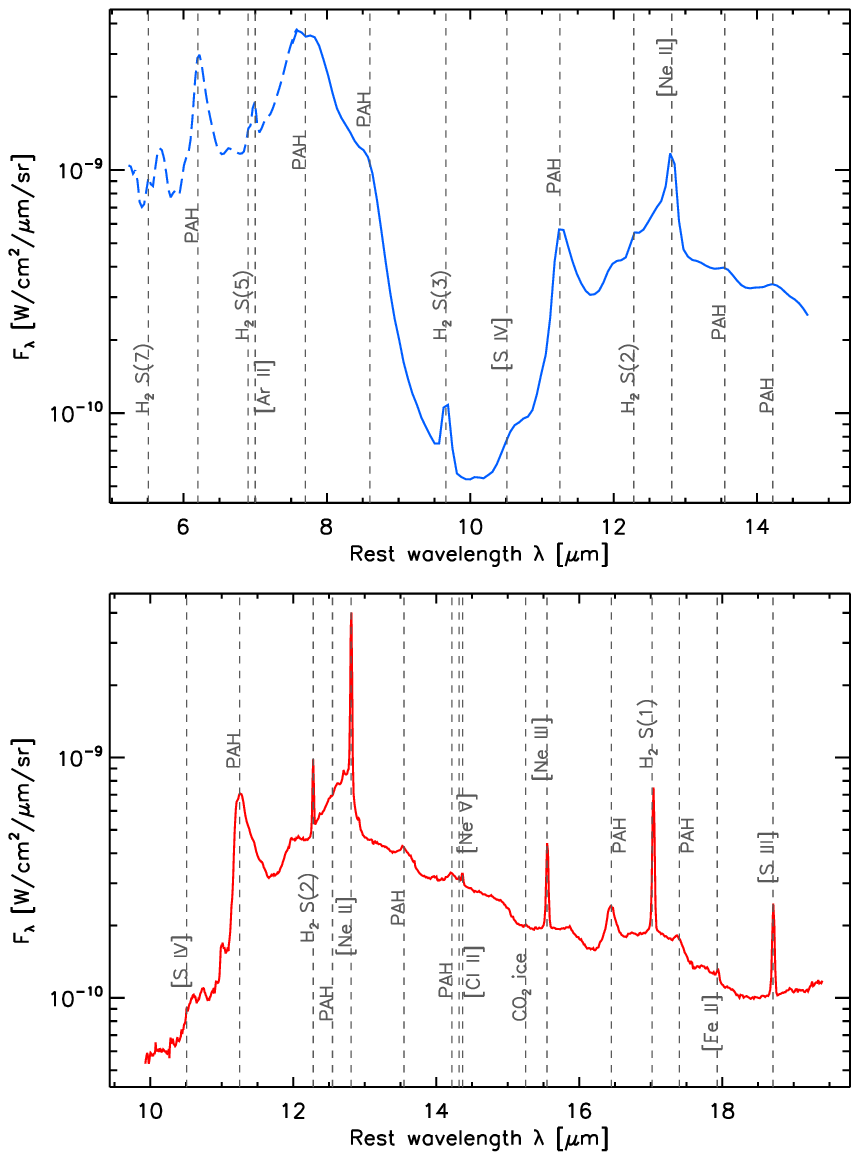}\hspace*{\fill}

   
  \caption{\footnotesize{\textit{Top panel} - Average spectrum of the equivalent field of view of the SH map of the nuclear region of NGC~4945, as seen in the SL map. The SL2 order is shown with a \textit{dashed} line. The most significant fine structure emission lines and PAH features in the SL wave range are labelled, in particular the ones that overlap with the SH waverange. \textit{Bottom panel} - Average spectrum of the whole field of view of the SH map. The higher spectral resolution of the IRS/SH module allows to distinguish emission lines like \nev14.32\mum\ and \clii14.37\mum, otherwise blended in the IRS/SL module. These spectra are not corrected for extinction.}}
  \label{fig:average-FOV-spectra}
\end{figure}

The average spectrum, considering the whole field of view (FOV) of the IRS/SH map, and the average spectrum of the IRS/SL map (extracted from a similar FOV as that of the SH map) are shown in the \textit{top} and \textit{bottom} panel of Fig.~\ref{fig:average-FOV-spectra}, respectively. The most important fine-structure emission lines and PAH features are indicated in both spectra.

We developed our own IDL procedures to work on the 1-D spectra that were used to build the final maps. 
These 1-D spectra were extracted from the clean SH cubes by averaging the spectra over a moving 2$\times$2 pixel aperture. The whole field of view was covered with a step size of 1 pixel, so the reduced map contains only 10$\times$10 pixels of $\sim2.3''$ each (where $1''$ corresponds to $\approx18.5~\rm pc$ at the distance of $3.82~\rm Mpc$) but covering the same area as the original map. However, the last slit observation of the IRS/SH map covers just about one third of the spatial region associated with each pixel. Therefore, we conclude that the last column of the original 11$\times$11 grid map produced by CUBISM is not representative of the actual spatial region associated with that slit column, and we do not use it in our maps. 
So the final maps reduced with the 2$\times$2 aperture results in 9$\times$9 pixel maps, covering a slightly smaller region than the original maps.
The coordinates assigned to each pixel of the resulting map are those of the center of the 2$\times$2 aperture, and the pixel fluxes reported in Figures \ref{fig:SH-maps1}, \ref{fig:SH-maps2}, \ref{fig:SH-maps1-extcorr} and \ref{fig:SH-maps2-extcorr} are the average fluxes in the 2$\times$2 apertures.
Using the 2$\times$2 pixel extraction box the S/N of the spectra is increased by a factor 2 (from the $\sqrt{4}$ pixels), and reduces the effect of the mismatch between orders mentioned above. 
This corresponds to a standard procedure recommended to obtain IRS maps of extended sources.

The average Full Width Half maximum (FWHM) of the IRS/SH Point Spread Function (PSF) has been estimated to be $\sim5.16''$ and $\sim4.31''$ for the direction along the spectral dispersion axis and the spatial axis of the slit, respectively, while the corresponding average PSF FWHM values of the SL2 (mostly used in this work) are $\sim3.76''$ and $3.27''$, respectively \citep[their appendix A]{pereira10}. The $1$-$\sigma$ deviation around the average centroid position in SL modules was found to be $\sim20$ times smaller than the SL pixel size ($1.85''$). While the maximum $1$-$\sigma$ deviation in the SH module was found to be $\sim0.6''$, which is considered small compared to the SH pixel size ($2.26''$). 

With the aim to determine whether our line emission maps are resolved, we retrieved a partial map of 3$\times$3 SH slit pointings of the star P Cygni (AOR 13049088). The map consists of $5''$ steps parallel to the slit and $2''$ steps perpendicular to the slit, similar to the map spacing we have for NGC~4945. These pointings lead to a 10$\times$3 pixel map. In order to test the continuum drop off along the (10 pixels) slit, 2$\times$2 aperture spectra were extracted in the same way as we do for the SH maps of NGC~4945. Then the continuum flux at 15~\mum\ was measured in each position, resulting in a mostly Gaussian variation of the flux along the slit. 
The FWHM ranges from 6 to 7 arcsec between 10 and 19 microns, far less than the ratio 19/10 expected over this range for the PSF of a diffraction limited telescope.
The same procedure was applied to the 12$\times$6 map of the star KsiDra used by \citet{pereira10} (AORs 16294912 and 16340224). In this case, a FWHM in the range $5.8''-6.4''$ was found in both the parallel and the perpendicular (to the slit) directions.

From the SH 2$\times$2 aperture average spectra we obtained maps of the starburst tracers \siv\ at 10.51 \mum, \neii\ at 12.81 \mum, \clii\ at 14.37 \mum, \neiii\ at 15.56 \mum, \siii\ at 18.71 \mum, as well as the AGN tracer \nev\ at 14.32 \mum, and the pure rotational molecular hydrogen lines, H$_2$ S(2) and H$_2$ S(1) at 12.3 \mum\ and 17.0 \mum, respectively. For each line we first estimate a continuum level by fitting a second-order polynomial to the base of the emission line, or a cubic spline with selected anchor points (pivots) when the emission line was on top of a complex PAH or silicate feature. 
We applied the same aperture reduction procedure for the SL map, from where we obtained a map of the silicate absorption feature at 9.7~\mum\ and the corresponding silicate strength as described in Sec.~\ref{sec:silicate}. We also obtained a map of the the molecular hydrogen line H$_2$ S(3) at 9.66~\mum.

\begin{figure}[tp]
  \centering
  \includegraphics[width=8cm,angle=0]{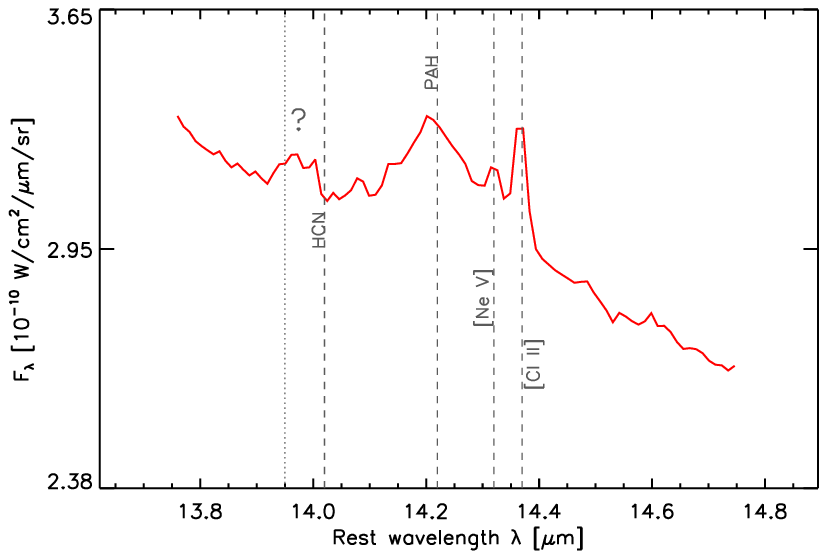}
  \caption{\footnotesize{Zoom into the \nev\ complex wavelength range from the average IRS/SH spectrum of Fig.~\ref{fig:average-FOV-spectra}. At these wavelengths we find the first bending mode of \hcn\ at $14.02$~\mum, and the AGN tracer \nev\ at $14.22$~\mum, almost blended with the fine-structure emission line \clii\ at $14.37$~\mum, on the slope of the PAH feature around $14.22$~\mum. The dotted line shows the wavelength at which we found an order mismatch, probably producing the unknown feature observed next to the \hcn~$14.02$~\mum, which could be affected.}}
  \label{fig:average-FOV-NeV}
\end{figure}

\subsection{The spectral complex around \nev~14.32~\mum}\label{sec:NeV-complex}

The complexity of the spectral range around the \nev~14.32~\mum\ is shown in Fig.~\ref{fig:average-FOV-NeV}. An unknown feature at $\sim$13.95~\mum\ seems to be present just next to the first bending mode of \hcn\ at $14.02$~\mum.
Depending on the temperature of the surrounding molecular gas, \hcn\ can be vibrationally excited by absorbing the infrared photons at $14.0$~\mum. This will produce a subsequent cascade process that can enhance \hcn\ rotational lines in the (sub-)millimeter range. This corresponds to the IR-pumping scenario proposed to explain the bright \hcn\ $J=1\rightarrow 0$ (and higher) transition observed in some ULIRGs and Galactic star-forming regions \citep[e.g.,][]{aalto95, garcia06, guelin07, aalto07a, pb10}. Although, due to the order mismatch (mentioned in Sec.~\ref{sec:results}) observed between $\sim13.90$ \mum\ and $\sim14$ \mum, also seen by \citet{pereira10}, we are unable to conclude whether we detect or not the \hcn~$14.02$~\mum\ feature, either in emission or in absorption.

A larger PAH feature is observed around $\sim$14.22~\mum, and on its rightmost (and less steep) slope the fine-structure emission lines of \nev~14.32~\mum\ and \clii~14.37~\mum\ are found. 
Chlorine, which has an ionization potential of $12.97~\rm eV$, is known to play an important role in characterizing the neutral gas components in the ISM. When H$_2$ is abundant (optically thick in the FUV), it reacts (exothermically by $0.17~\rm eV$) with \clii\ to form HCl$^+$, which leads to the formation of \cli\ and \hi\ \citep{jura74, jura78}. This means, chlorine is predominantly ionized in \hi\ regions while it is predominantly neutral when cold H$_2$ components are present \citep{sonnentrucker02, sonnentrucker03}. The \clii~14.37~\mum\ fine-structure emission line was clearly detected ($>$3$\sigma$) in the whole region mapped with the IRS/SH module, and presents a similar (although more spreadout) distribution than that of the \neii~12.81~\mum\ line.

Because its ionization potential ($97.1~\rm eV$) is too large to allow production by main-sequence stars, the fine-structure line \nev~14.32~\mum\ is commonly used to probe the narrow line region of AGNs \citep[e.g.,][and references therein]{moorwood96b, genzel98, sturm02, armus07, alonsoh09, baum10, willett10}. It has recently been used as a diagnostic tool to unambiguously identify AGN in galaxies that have not been identified as such using optical spectroscopy \citep{goulding09}.

\subsection{Estimating the line emission flux}\label{sec:flux-estimate}

\begin{figure}[!tp]
  \centering
  \includegraphics[width=8cm,angle=0]{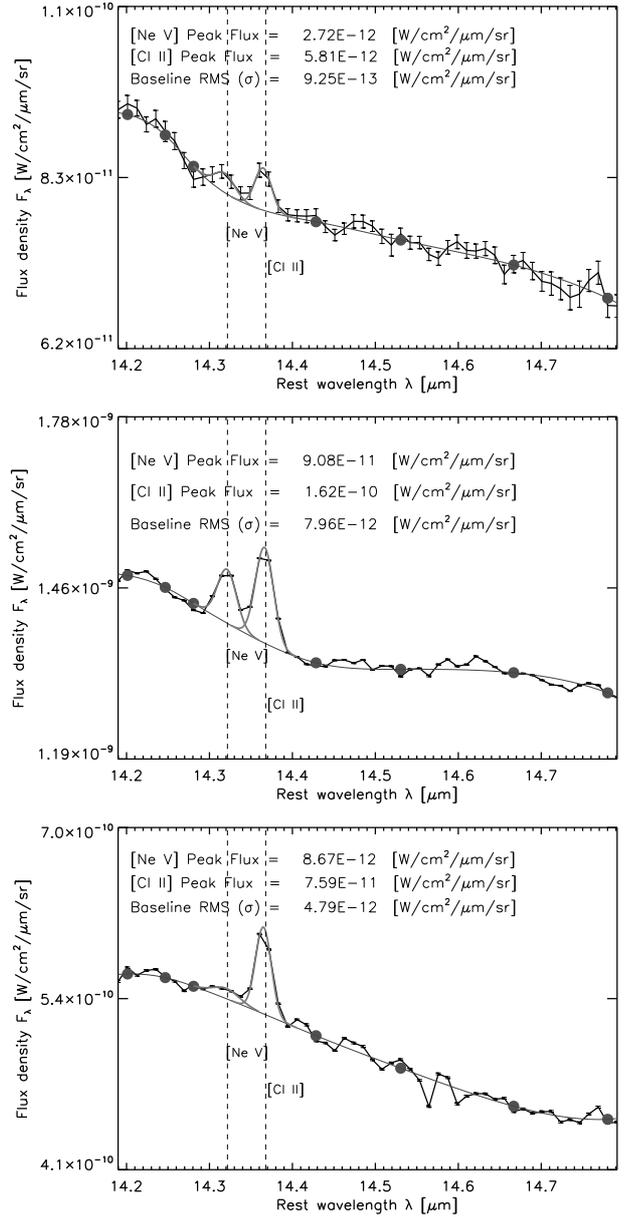}
  \caption{\footnotesize{Gaussian fit of the \nev~14.32~\mum\ and \clii~14.37~\mum\ fine-structure emission lines at three different pixels of the IRS/SH map. The error bars indicate the uncertainties of the spectral data. The filled circles indicates the pivots used for the cubic spline interpolation of the baseline. The \textit{top panel} shows less than 3$\sigma$ detection for \nev\ but a clear detection of the \clii\ line. The spectrum of the \textit{middle panel} shows clear detections of both lines. The \textit{bottom panel} shows $<$3$\sigma$ detection for \nev, in spite of the low uncertainties in the data points.
  }}
  \label{fig:Gaussian-fit-NeV}
\end{figure}

The emission fluxes were estimated by fitting a Gaussian profile to the fine-structure lines and integrating the flux above a local continuum, which was estimated from a cubic spline interpolation of selected anchor points (pivots). 
Since we we want to determine the actual spatial distribution of the emission lines in the nuclear region of NGC~4945,
we set a strict 3$\sigma$ level detection for all the fine-structure emission lines, and at least two data points (or a full width of $\sim$560~\kms at 14.32~\mum) in the profile to consider a feature as a real emission line in the spectrum. For a robust estimate of the local continuum we adopted an average flux of five points around each of the selected pivots used in the cubic spline interpolation.
Figure~\ref{fig:Gaussian-fit-NeV} shows the Gaussian fit of the \nev~14.32~\mum\ and \clii~14.37~\mum\ fine-structure emission lines at three different positions of the IRS/SH map. The uncertainty of each spectral data point is indicated by the error bars. The pivots used for the cubic spline interpolation of the local continuum (baseline) are shown with filled circles. These pivots are at the wavelengths 14.201~\mum, 14.247~\mum, 14.281~\mum, 14.428~\mum, 14.531~\mum, 14.667~\mum, and 14.780~\mum.

\begin{figure*}[!ht]
  


   
  \hfill\includegraphics[width=14cm]{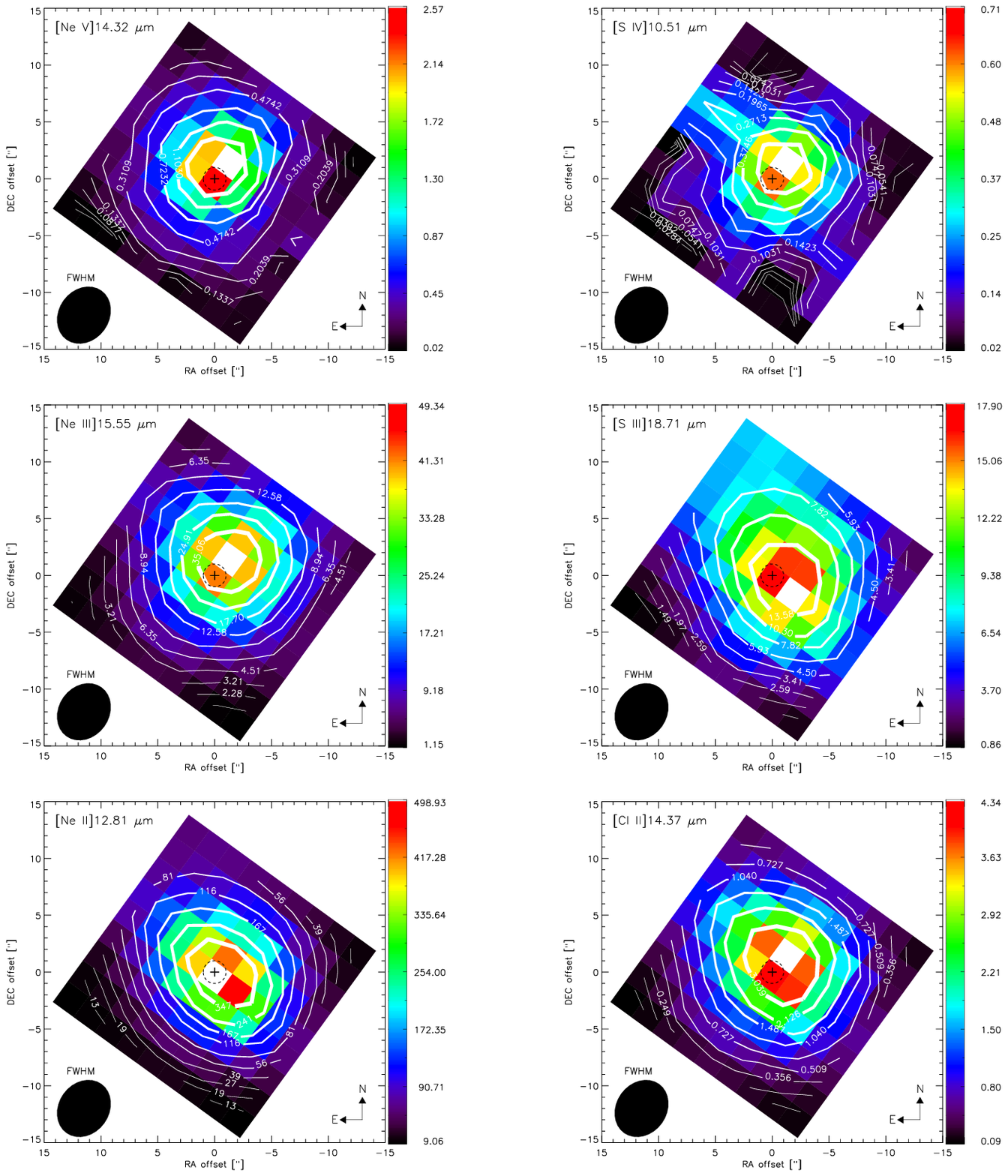}\hspace*{\fill}

  \caption{\footnotesize{IRS/SH surface brightness ($10^{-12}~\Wcmsr$) maps of the fine-structure lines (\textit{left panels}, from top to bottom) \nev~14.32~\mum, \neiii~15.55~\mum, \neii~12.81~\mum, (\textit{right panels}, from top to bottom) \siv~10.51~\mum, \siii~18.71~\mum, and \clii~14.37~\mum. The peak fluxes are shown with a white pixel. These fluxes are not corrected for extinction, and each of the 9x9 pixels correspond to the fitted flux in the 2x2 aperture extracted spectrum as described in Sections~\ref{sec:data-reduction} and \ref{sec:flux-estimate}. The dashed-line circle represents the $\pm 1''$ pointing accuracy of Spitzer, and the contour lines are labelled. The reference ($\Delta\alpha=0,\Delta\delta=0$) is marked with a cross and corresponds to the position R.A.(J2000)=13:05:27.477, Dec.(J2000)=-49:28:05.57 of the H$_2$O mega maser reported by \citet{greenhill97}. }}
  \label{fig:SH-maps1}
\end{figure*}

\begin{figure*}[!ht]
  



  \hfill\includegraphics[width=14.2cm]{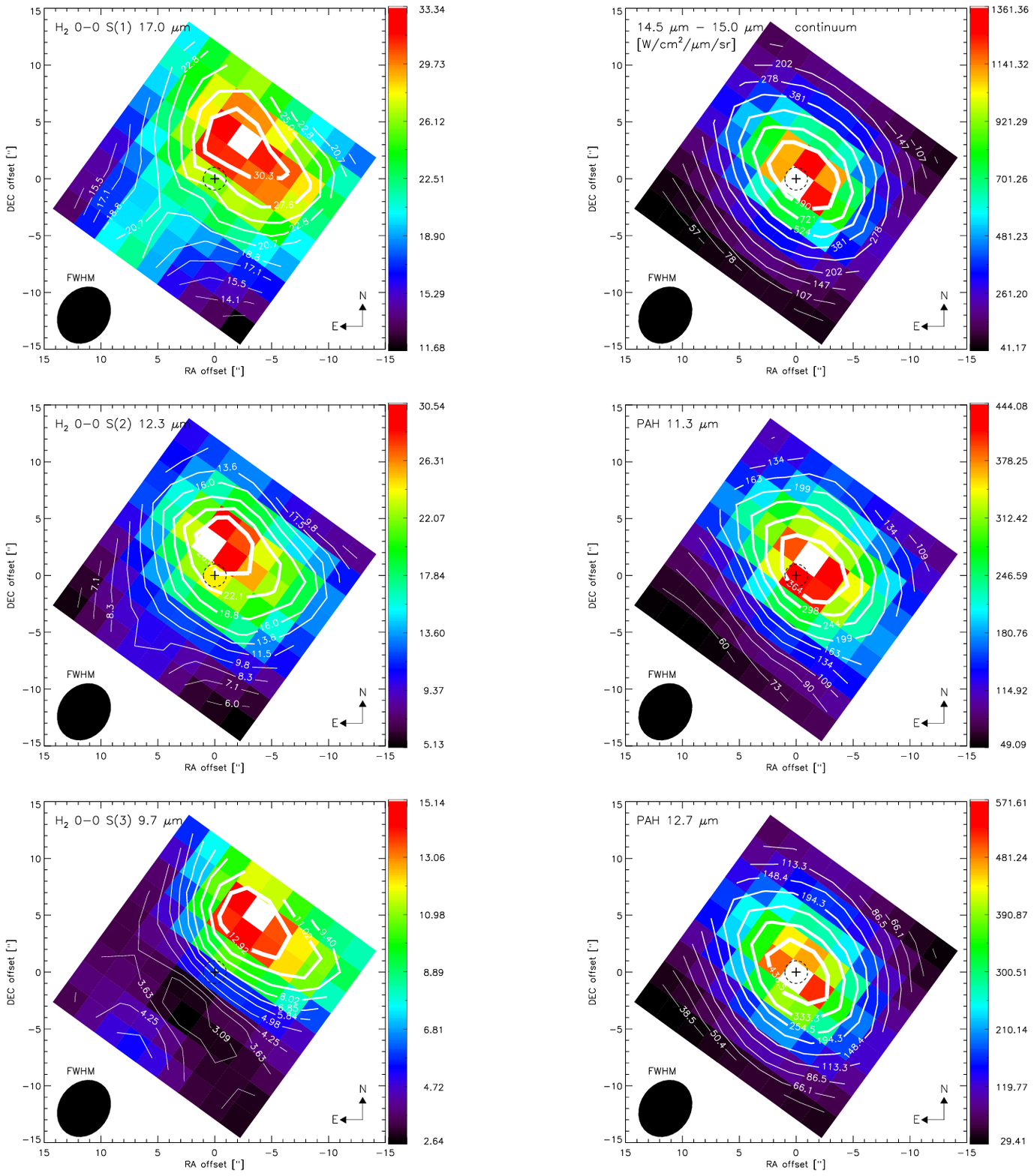}\hspace*{\fill}
  
  \caption{\footnotesize{IRS/SH surface brightness ($10^{-12}~\Wcmsr$) maps of the molecular hydrogen lines (\textit{left panels}, from top to bottom) H$_2$ S(1) and H$_2$ S(2) at 17.0 \mum\ and 12.3 \mum, respectively, and the H$_2$ S(3) line at 9.7 \mum\ from the IRS/SL map. The \textit{right panel} shows (from top to bottom) the average continuum flux density ($10^{-12}~\Wcmusr$) between 14.5~\mum\ and 15.0~\mum, and the surface brightness ($10^{-12}~\Wcmsr$) of the PAH features at 11.3~\mum\ and 12.7~\mum. The details in the maps are the same as in Figure~\ref{fig:SH-maps1}.}}
  \label{fig:SH-maps2}
\end{figure*}

The \textit{top panel} of Fig.~\ref{fig:Gaussian-fit-NeV} shows the spectrum in a region of the map with relatively high uncertainties ($\sim$1.4\% on average) in the spectral data and with high RMS (as computed from the baseline, and indicated in the legend of the plots) that leads to less than a 3$\sigma$ detection for \nev, but a clear detection ($>$3$\sigma$) for the \clii\ line. The spectrum of the \textit{middle panel} was extracted from the central 2$\times$2 pixel aperture of the map, which has a S/N ratio $\sim$9 in the line and $>$100 in the continuum, and is among the highest S/N in the whole IRS/SH map. This spectrum shows very low uncertainties ($\sim$0.06\% on average) in the data, and clear detections of both lines. 
Although with low uncertainties in the data ($\sim$0.13\%), the \textit{bottom panel} of Fig.~\ref{fig:Gaussian-fit-NeV} shows less than a 3$\sigma$ detection for \nev\ (with a S/N$\sim$1.8), but a strong detection of \clii~14.37~\mum.
With this Gaussian fit procedure, we produced 9$\times$9 pixels maps (as described above) of several fine-structure lines, PAH features and the silicate absorption feature at $9.7$~\mum\ (discussed in Sec.~\ref{sec:silicate}), covering a region of about $23''\times23''$ (about 426$\times$426$~\rm pc^2$). The SH maps are shown in Figs.~\ref{fig:SH-maps1} and \ref{fig:SH-maps2}. In the case of the \nev~14.32~\mum\ the pixels with surface brightness lower than $0.1\times10^{-12}~\Wcmsr$ correspond to a detection level $<3\sigma$.

For the particular case of the \nev\ line, a Gaussian fit of the uncorrected for extinction map indicates that this emission is only marginally resolved along the major axis, given its FWHM of $6.4''$. Along the minor axis the FWHM is $7.3''$. After extinction correction (discussed in Secs.~\ref{sec:Av-HK} and \ref{sec:silicate}), the \nev\ emission appears unresolved. The other emission lines (uncorrected for extinction) are well resolved (FWHM$>7''$).

\subsection{Visual extinction}\label{sec:Av-HK}

Lower limits on the reddening in the nuclear region of NGC~4945 were obtained from the HST-NICMOS $H-K$ color image \citep{marconi00}. An average color $H-K = 1.1$, yielding an $A_{\rm V}\approx 11~\rm mag$, was observed in the region of the Pa$\alpha$ ring. This $A_{\rm V}$ is comparable with extinctions larger than $13~\rm mag$, as estimated from the Br$\alpha$/Br$\beta$ and Pa$\alpha$/H$\alpha$ ratios \citep{moorwood88, marconi00}.

In order to compare with our data, we produced an image of $A_{\rm V}$ at the same resolution of the IRS/SH maps, using the HST-NICMOS $H-K$ colour image. First, we use the same definition of foreground screen extinction given in \citet{marconi00}

\begin{equation}\label{eq:Av(H-K)}
A_{\rm V}(H-K) = \frac{E(H-K)}{c(H) - c(K)}, 
\end{equation}

\noindent
where $E(H-K)$ is the colour excess that can be obtained from the difference between the observed and intrinsic colours, $E(H-K) = (H-K)-(H-K)_0$. We use the average intrinsic colour $(H-K)_0\approx 0.22\pm0.1~\rm mag$ of spiral and elliptical galaxies reported by Hunt et al. (1997). The $c(H)$ and $c(K)$ coefficients are derived from the extinction law, $A_{\lambda}=c(\lambda)A_{\rm V}$. But instead of using the $A_\lambda\varpropto(\lambda/1~\mu m)^{-1.75}$ law ($\lambda>1~\mu m$) assumed by \citet{marconi00}, we used the extinction law for the local ISM reported by \citet{chiar06}, which considers solid and porous spheres and a continuous distribution of ellipsoids in the extinction profiles used for the amorphous silicates in the $9.7$~\mum\ absorption feature. We interpolated the $c$ coefficients at $\lambda=1.606~\mu m$ and $\lambda=2.218~\mu m$ of the HST-NICMOS $H$ and $K$ bands, respectively. 
In \citet{chiar06} the local ISM continuum extinction was found to be described by the expression
\begin{equation}\label{eq:ext-law}
 log(A_\lambda/A_K)=0.65-2.4log(\lambda)+1.34log(\lambda)^2,
\end{equation}
\noindent
using $A_K/A_{\rm V}=0.09$ to normalize the extinction to the $K$ band \citep{whittet03}. Beyond 8~\mum\ the silicate profile of WR-98A is superimposed using $A_{\rm V}/\tau(9.7~\mu m)=18$ \citep{roche84}. So we used eq.(\ref{eq:ext-law}) and $A_K/A_{\rm V}=0.09$ to estimate the $c(H)$ and $c(K)$ coefficients for the stellar light-based extinction $A_{\rm V}(H-K)$.

\begin{figure}[!tp]


  \hspace*{\fill}\includegraphics[width=6.5cm,angle=0]{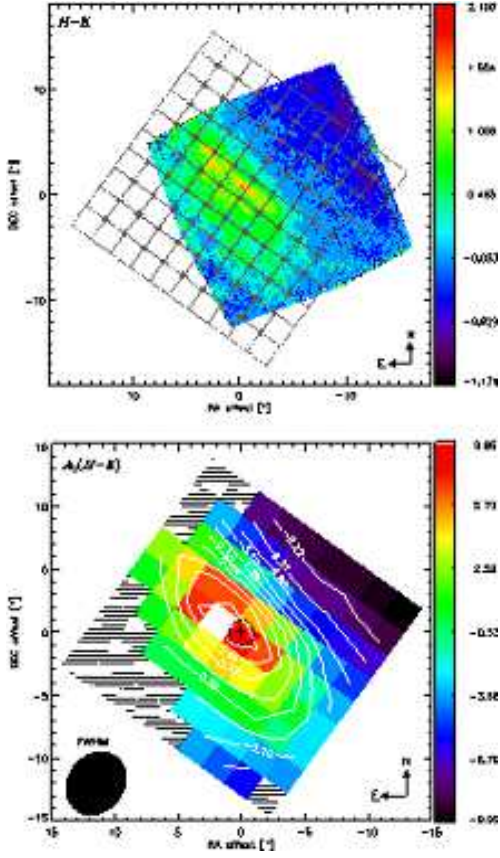}\hspace*{\fill}

  \caption{\footnotesize{\textit{Top panel} - HST-NICMOS $H-K$ colour image ($\rm mag$) reported by \citet{marconi00}. The Spitzer-IRS SH grid is overlaid to show the difference in resolution and the different orientation of the fields of view. \textit{Bottom panel} - Foreground screen extinction $A_{\rm V}(H-K)$ estimated from the $H-K$ colour image at the resolution of the IRS/SH map.
  }}
  \label{fig:H-K-image}
\end{figure}

The \textit{top panel} of Fig.~\ref{fig:H-K-image} shows the HST-NICMOS $H-K~\rm mag$ image reported by \citet{marconi00} with the Spitzer-IRS SH grid overlaid. We took all the pixels of the $H-K$ image (which have higher resolution than our maps) that fall within a 2$\times$2 pixel aperture of the SH grid map, and computed the average $A_{\rm V}$ for that particular pixel. After doing this for every single pixel in the SH grid we obtain the foreground extinction (at the resolution of the SH map) shown in the \textit{bottom panel} of Fig.~\ref{fig:H-K-image}. 
The SH pixels that cover less than 400 pixels in the $H-K$ image are shown in dashed lines.
The peak of this stellar light-derived extinction lies about $2.3''$ (one pixel) northeast of the H$_2$O mega maser.

\subsection{The deep silicate absorption around 9.7~\mum}\label{sec:silicate}

From the IRS/SL cubes we can study the amorphous silicate grains, which present a broad and deep absorption feature around 9.7~\mum. The presence of PAH emission along the same line of sight as the silicate absorption makes it hard to measure the apparent silicate depth. If one assumes the PAH emission to be foreground to the silicate absorption, then a different apparent silicate depth is found than when it is assumed that they are mixed. Silicate absorption can also be foreground to the PAH emission. Here we use the method proposed by \citet[their Fig.2]{spoon07} for absorption-dominated spectra to infer the apparent strength of this silicate feature.
First, we extracted 1-D spectra from the SL map rebinned to the 2$\times$2 pixel aperture of the SH map described in Sec.~\ref{sec:results}. For each spectrum, we adopted a local mid-infrared continuum by interpolating the feature-free continuum pivots at 5.3~\mum\ and 13.55~\mum.

Because the S/N is not uniform for all the pixels of the SL map, and the features around 5.3~\mum\ change depending on the proximity to the nucleus of NGC~4945, we used a similar procedure as the one used for \nev~14.32~\mum\ (Sec.~\ref{sec:NeV-complex}). The flux assigned to the pivot at 5.3~\mum\ was adopted as the minimum flux observed between 5.1~\mum\ and 5.5~\mum. In order to avoid contamination from the H$_2$ S(3) 9.66~\mum\ molecular line, the flux estimated for the deepest point adopted at 9.85~\mum\ was the median flux between 9.7~\mum\ and 10.0~\mum. The flux estimated for the pivot at 13.55~\mum\ was the median flux observed between 12.9~\mum\ and 14.0~\mum.
Then we computed the ratio between the observed flux ($f_{\rm obs}$) and the estimated continuum flux ($f_{\rm cont}$) at 9.85~\mum, and we obtained the silicate strength $S_{\rm sil}$ as

\begin{equation}
 S_{\rm sil}=ln\left[ \frac{f_{\rm obs}(9.85~\mu \rm m)}{f_{\rm cont}(9.85~\mu \rm m)} \right].
\end{equation}

\begin{figure}[!tp]


  \hspace*{\fill}\includegraphics[width=6.5cm,angle=0]{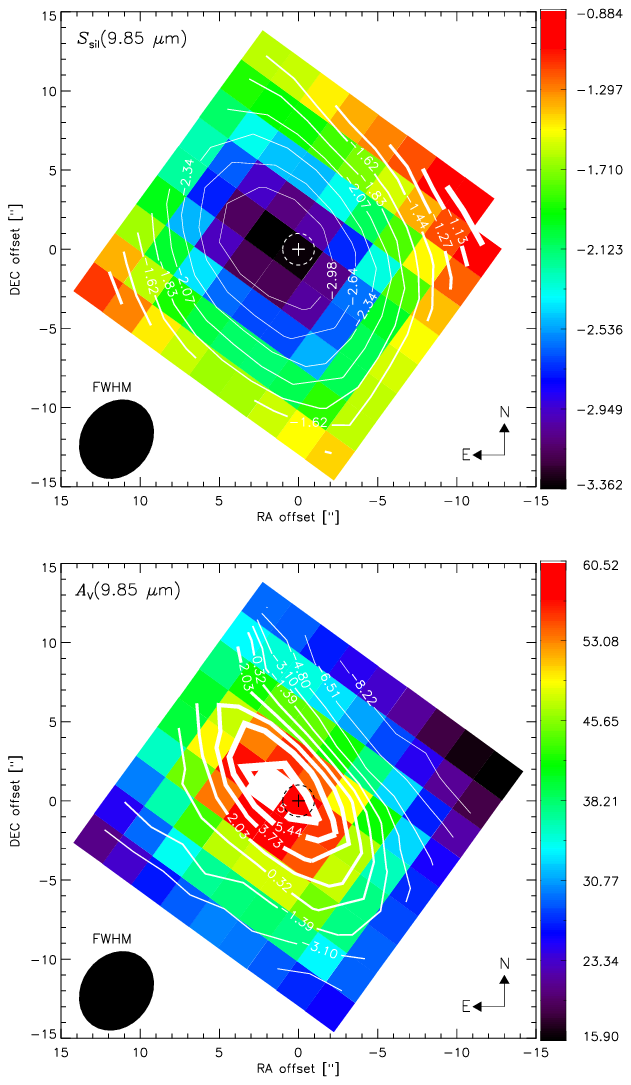}\hspace*{\fill}

  \caption{\footnotesize{\textit{Top panel} - IRS/SL map of the apparent silicate strength $S_{\rm sil}$ estimated at 9.85~\mum. \textit{Bottom panel} - Silicate-based extinction $A_{\rm V}(9.85~\mu \rm m)$ map estimated from the average visual extinction to silicate optical depth ratio $A_{\rm V}/\tau(9.7~\mu m)=18$ for the local ISM \citep{roche84, rieke85} and assuming $\tau(9.7~\mu m)\approx S_{\rm sil}(9.85~\mu \rm m)$. The contour lines of the stellar light-based extinction $A_{\rm V}(H-K)$ of Fig.~\ref{fig:H-K-image} are overlaid on the $A_{\rm V}(9.85~\mu \rm m)$ map (colour pixels).
  }}
  \label{fig:Tau-sil}
\end{figure}

For sources with a silicate absorption feature, $S_{\rm sil}$ can be interpreted as the negative of the apparent silicate optical depth ($\tau_{9.7~\mu \rm m}$).
The \textit{top panel} of Fig.~\ref{fig:Tau-sil} shows the IRS/SL map (rebinned to the 2$\times$2 aperture of the IRS/SH map) of the apparent silicate strength at 9.85~\mum. The silicate-based extinction $A_{\rm V}(9.85~\mu \rm m)$, shown in the \textit{bottom panel} of Fig.~\ref{fig:Tau-sil}, was estimated from the average visual extinction to silicate optical depth ratio $A_{\rm V}/\tau(9.7~\mu \rm m)=18$, which is appropriate for the local ISM \citep{roche84, rieke85}. The spatial distribution of the silicate-based extinction is similar to that of the stellar light-based extinction $A_{\rm V}(H-K)$ estimated in Sec.\ref{sec:Av-HK}. The peak extinction is found at the same relative position ($\Delta~\rm R.A.\approx2''$, $\Delta~\rm Dec\approx1''$), at about $2.3''$ (one pixel) northeast of the H$_2$O mega maser \citep{greenhill97}. 

It is known that extinction estimated from optical or near-infrared observations generally underestimates the actual extinction if the environment probed is optically thick at the emission lines observed. If most of the emitting region is obscured, as in the case of the nucleus of NGC~4945, the visual extinction estimate is representative of the surface of the obscured region and not the region itself. This effect is reflected in the different extinctions derived from the $H-K$ optical image and the silicate-based estimate derived from our mid-IR observations, where the peak extinction is about 7 times stronger than that of the stellar light-based estimate. Our $A_{\rm V}(9.85~\mu \rm m)$ is also a factor $\sim$1.7 higher than the extinction ($A_{\rm V}=36^{+18}_{-11}~\rm mag$) previously inferred from ISO observations of the \siii~18.7/33.5~\mum\ line ratio \citep{spoon00}. This difference can be explained by the peaked nature of the silicate-based extinction map and because of the larger apertures ($\ge14''\times20''$) of the ISO observations, which averages out the extinction to the lower value.

\begin{figure*}[!htp]



  \hfill\includegraphics[width=14.2cm]{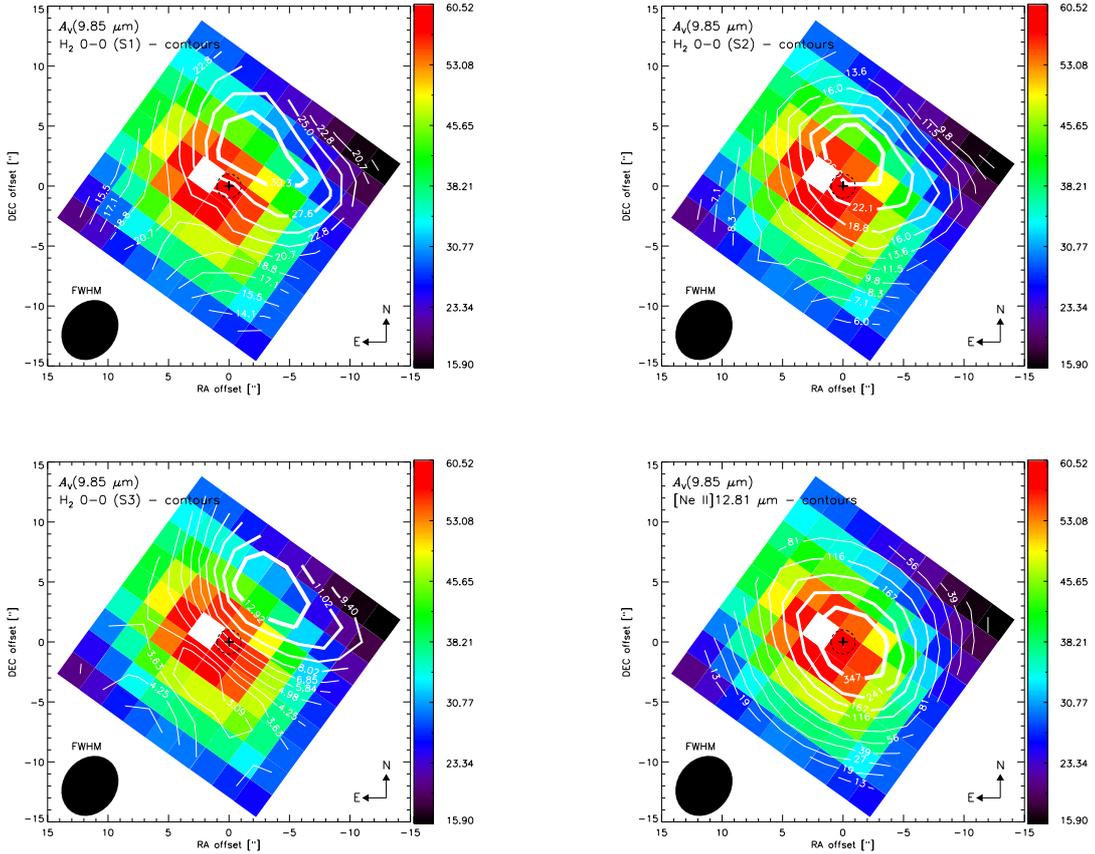}\hspace*{\fill}

  \caption{\footnotesize{IRS/SL map of the silicate-based extinction $A_{\rm V}(9.85~\mu \rm m)$ with the contour lines of the IRS/SH map of the surface brightness (in units of $10^{-12}~\Wcmsr$) of the molecular hydrogen line H$_2$ S(1) 17.0 \mum\ (\textit{top left panel}), the IRS/SH map of the H$_2$ S(2) 12.3 \mum\ line (\textit{top right panel}), the IRS/SL map of the H$_2$ S(3) 9.7 \mum\ line (\textit{bottom left panel}), and the IRS/SH map of the \neii~12.81 \mum\ (\textit{bottom right panel}). The figures show that the H$_2$ emission is stronger in a region $\gtrsim$2.3$''$ away from the peak obscuration, although the H$_2$ S(2) is the closest to the peak obscuration and to the H$_2$O mega maser, as well as the \neii, which peaks at the H$_2$O mega maser (within $\sim1''$), as seen in Fig.~\ref{fig:SH-maps1}.
  }}
  \label{fig:AvSil-H2}
\end{figure*}

Figure~\ref{fig:AvSil-H2} shows the contour lines of the IRS/SH surface brightness map of the molecular hydrogen line H$_2$ S(1) 17.0 \mum\ (\textit{top left}), H$_2$ S(2) 12.3 \mum\ (\textit{top right}), the IRS/SL map of the H$_2$ S(3) 9.7 \mum\ line (\textit{bottom left}), and the \neii~12.81 \mum\ line (\textit{bottom right}), overlaid on the silicate-based extinction $A_{\rm V}(9.85~\mu \rm m)$ map. The molecular hydrogen emission avoids the obscured nucleus and peaks between one and three pixels away from the highest obscuration. Although the H$_2$ S(2) 12.3 \mum\ (Fig.~\ref{fig:SH-maps2}) has a similar distribution as the H$_2$ S(1) line, its peak emission lies closer ($\sim2.3''$, one pixel) to the peak obscuration and to the H$_2$O mega maser than the other H$_2$ lines.


\begin{figure*}[!htp]




  \hfill\includegraphics[width=14.2cm]{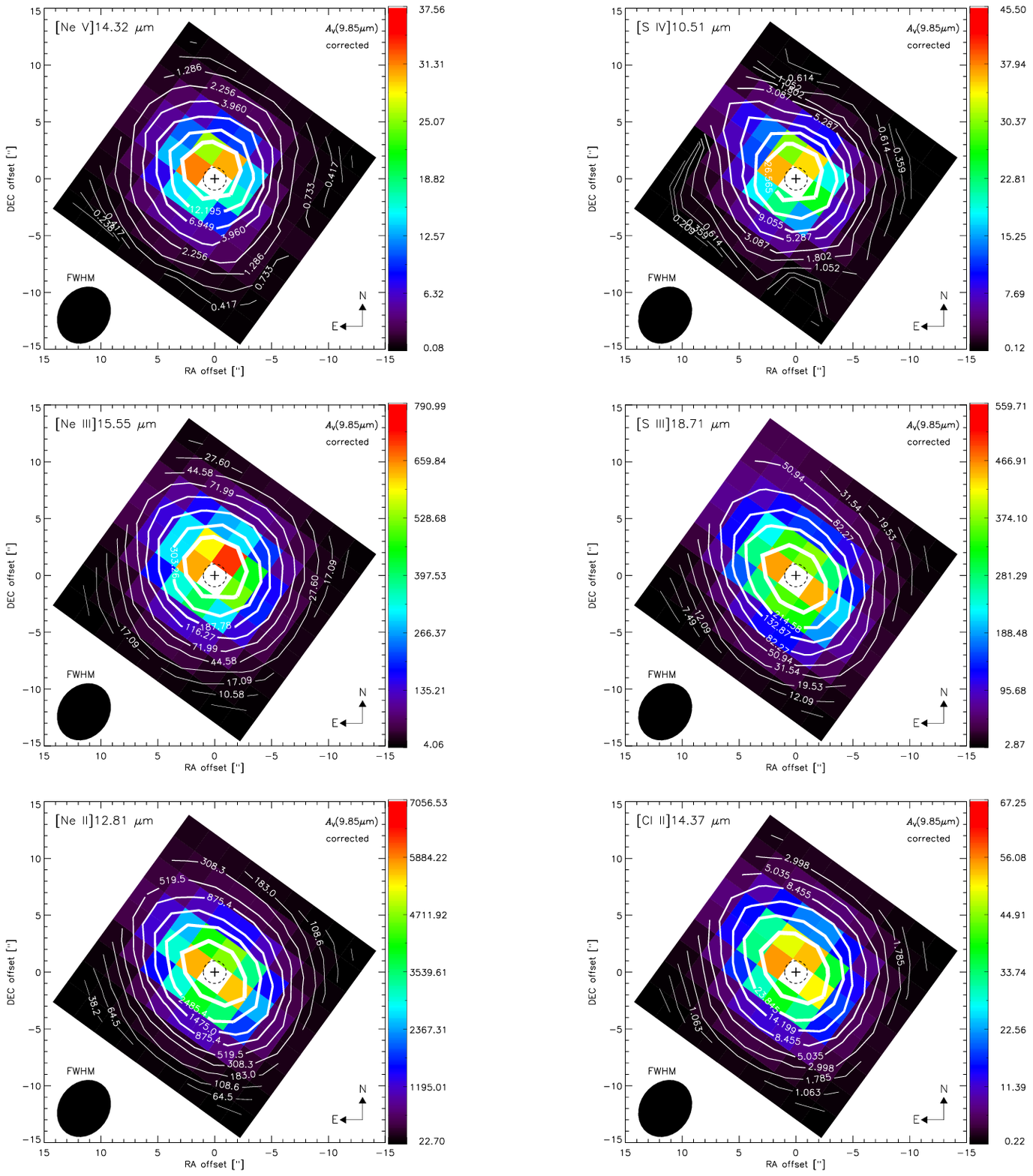}\hspace*{\fill}
   
  \caption{\footnotesize{IRS/SH surface brightness ($10^{-12}~\Wcmsr$) maps, corrected for extinction using the extinction law for the local ISM from \citet{chiar06} and the silicate-based extinction $A_{\rm V}(9.85~\mu \rm m)$ (Figs.\ref{fig:Tau-sil} and \ref{fig:AvSil-H2}). The \textit{left panels} show (from top to bottom) the fine-structure lines \nev14.32~\mum, \neiii15.55~\mum, and \neii12.81~\mum. The \textit{right panels} show (from top to bottom) the \siv10.51~\mum, \siii18.71~\mum, and \clii14.37~\mum\ lines. The extinction corrected fluxes and notations are as in Fig.~\ref{fig:SH-maps1}. Note that after correcting for extinction, all these emission lines (and the average continuum) peak at about the same position, that of the H$_2$O mega maser.}}
  \label{fig:SH-maps1-extcorr}
\end{figure*}

\begin{figure*}[!htp]
  



  \hfill\includegraphics[width=14.2cm]{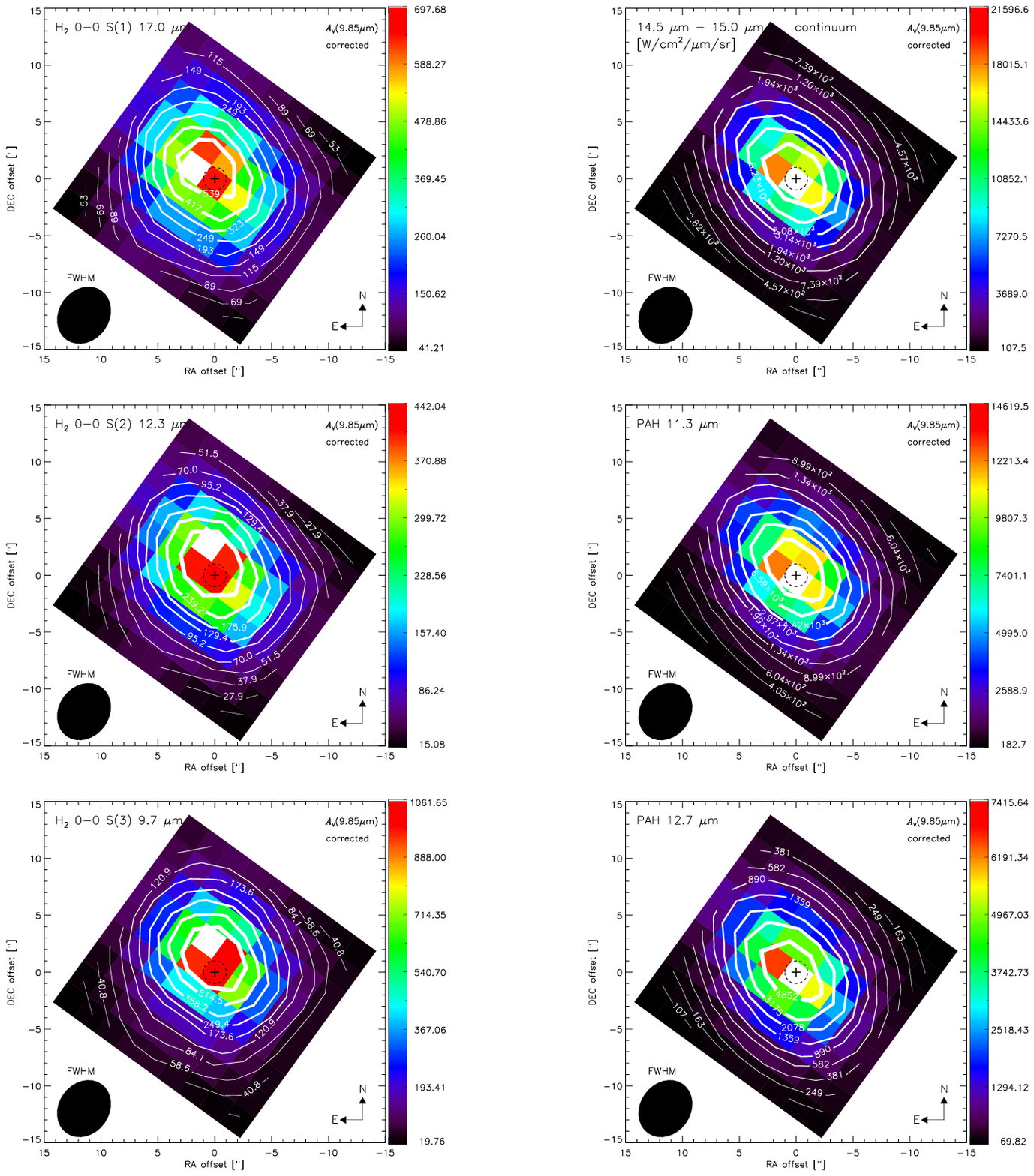}\hspace*{\fill}

  \caption{\footnotesize{IRS/SH surface brightness ($10^{-12}~\Wcmsr$) maps, corrected for extinction using the extinction law for the local ISM from \citet{chiar06} and the silicate-based extinction $A_{\rm V}(9.85~\mu \rm m)$ (Figs.\ref{fig:Tau-sil} and \ref{fig:AvSil-H2}). The molecular hydrogen lines H$_2$ S(1) 17.0 \mum, H$_2$ S(2) 12.3 \mum, and H$_2$ S(3) 9.7 \mum\ (from the IRS/SL map) are shown in the \textit{left panel}. The \textit{right panel} shows (from top to bottom) the average continuum flux density ($10^{-12}~\Wcmusr$) between 14.5~\mum\ and 15.0~\mum, and the the surface brightness ($10^{-12}~\Wcmsr$) of the PAH features at 11.3~\mum\ and 12.7~\mum. The extinction corrected fluxes and notations are as in Fig.~\ref{fig:SH-maps1}. After correcting for extinction, the \clii\ line and the PAH features also peak at about the same position of the H$_2$O mega maser.}}
  \label{fig:SH-maps2-extcorr}
\end{figure*}

Since the distribution of the silicate-based extinction $A_{\rm V}(9.85~\mu \rm m)$ is similar to that of the stellar light-based extinction $A_{\rm V}(H-K)$ we think that the mid-IR derived extinction towards the starburst ring may be accurate to correct the fluxes of the starburst tracers. That is, the elongated shape of the silicate map indicates that the obscuration is mostly associated with the starburst ring rather than with the line of sight to the AGN.
Hence, the $A_{\rm V}(9.85~\mu \rm m)$ map can be used to perform an extinction correction for species whose emission emerge from the disk. Species associated with the AGN (BLR and NLR gas) may be suffering more extinction than the disk does. 
Therefore, the corrected flux derived for the \nev~14.32~\mum\ corresponds only to the best lower limit we can derive from the current data. Figures~\ref{fig:SH-maps1-extcorr} and \ref{fig:SH-maps2-extcorr} show the same set of maps as in Figs.~\ref{fig:SH-maps1} and \ref{fig:SH-maps2}, but corrected for extinction using the estimated $A_{\rm V}(9.85~\mu \rm m)$ map. Note that after correcting for extinction, all the fine-structure emission lines, the average 14.5--15.0~\mum\ continuum, and the PAH features peak at about the same position, that of the H$_2$O mega maser. The corrected for extinction H$_2$ lines (Fig.~\ref{fig:SH-maps2-extcorr}) show an offset of $\sim2.3''$ (one pixel) with respect to the water maser. However, there is a difference of only $<1''$ between the centroids obtained from a two-dimensional Gaussian profile fit of all the lines.

   \begin{table}[htp]
      \caption[]{\footnotesize{LINE FLUXES$^{\mathrm{a}}$ FROM THE 10$\times$10 PIXEL ($\sim$426$\times$426 $\rm pc^2$) APERTURE IRS/SH MAP.}}
         \label{tab-c4:fluxes-fov}
         \centering
         \tabcolsep 5.8pt
         \scriptsize
         \begin{tabular}{lccccc}
            \hline\hline
	    \noalign{\smallskip}
            Line   &    $\lambda_0^{~\mathrm{b}}$   &   Flux$^{~\mathrm{c}}$   &   Flux($A_{\rm V}$)$^{~\mathrm{d}}$   &   FWHM$^{~\mathrm{e}}$   & \\
                   &    [$\mu \rm m$] & [$10^{-21}~\Wcm$] &     [$10^{-21}~\Wcm$]    &   [\kms] \\
            \noalign{\smallskip}
            \hline
            \noalign{\smallskip}

            \multirow{2}{*}{[S IV]}           & 10.5105 & 1.61  & 22.03  & 382.68        \\
		                            &       & 0.08  &  1.12  &  30.54         \\

	    \noalign{}\\
            \multirow{2}{*}{H$_2$ 0-0 S(2)} & 12.2786 & 154.07  &  889.94  &  525.66        \\
		                            &       &   1.99  &    11.57  &    6.23        \\

	    \noalign{}\\
            \multirow{2}{*}{[Ne II]}          & 12.8135  &  1201.08  &  6026.69  &  615.54     \\
		                            &        &   118.85  &   563.35  &   32.37     \\
            
	    \noalign{}\\
            \multirow{2}{*}{[Ne V]}           & 14.3217   &  5.34   & 27.93  &  584.70        \\
		                            &         &  0.53   &  0.64   &  15.44         \\

	    \noalign{}\\
            \multirow{2}{*}{[Cl II]}          & 14.3678  &  12.82  & 66.75  &  502.97   \\
		                            &        &   0.10  &  0.53   &   4.04    \\

	    \noalign{}\\
            \multirow{2}{*}{[Ne III]}          & 15.5551 &  127.01 & 741.49  &  640.81   \\
		                             &       &    2.79 &   18.03  &   12.74   \\

	    \noalign{}\\
            \multirow{2}{*}{H$_2$ 0-0 S(1)}  & 17.0348 & 252.60  & 1781.25  &  549.60 \\
		                             &       &   7.64  &  63.90  &   18.12  \\

	    \noalign{}\\
            \multirow{2}{*}{[S III]}           & 18.7129 & 74.87  &  618.00  &  562.71   \\
		                             &       &  1.37  &  10.90    &   9.59    \\

            \noalign{\smallskip}
            \hline	    		    

         \end{tabular}
\begin{list}{}{}
\scriptsize
\item[$^{\mathrm{a}}$] These are the fluxes obtained from the co-added spectra of the whole 10$\times$10 aperture of the SH map described in Sec.~\ref{sec:results}.
\item[$^{\mathrm{b}}$] Rest wavelength of the lines.
\item[$^{\mathrm{c}}$] The flux densities of each line are given in the first row. The row below shows the corresponding uncertainties.
\item[$^{\mathrm{d}}$] Flux density corrected for extinction using an average extinction value $A_{\rm V}\sim36$ mag estimated from the silicate-based $A_{\rm V}(9.85\mu \rm m)$ (Sec.~\ref{sec:silicate}). For [Ne V] this extinction correction may be insufficient. See Sec.~\ref{sec:NeV-dominated}.
\item[$^{\mathrm{e}}$] Line width in $\kms$ obtained from the Gaussian fit. The line profile is not corrected for the instrumental profile, which has a FWHM of 500 km/s.
\end{list}
\end{table}

The integrated flux densities of the co-added spectra from the whole 10$\times$10 aperture of the IRS/SH map (described in Sec.~\ref{sec:results}) are summarized in Table~\ref{tab-c4:fluxes-fov}. In this work we include only the fluxes of the most prominent emission lines. These fluxes are larger than those reported by \citet{bernards09}, because our 10$\times$10 aperture is larger than the SH staring aperture.
In Table~\ref{tab-c4:fluxes-fov} we also include the fluxes corrected with the silicate-based extinction $A_{\rm V}(9.85~\mu \rm m)$, and the line widths (FWHM) as estimated from the Gaussian fit.

\subsection{Rotation in the nuclear region}\label{sec:rotation-maps}

Even though the spectral resolution (R$\sim$600) of the IRS/SH module is relatively low in comparison to the resolution used in most kinematic studies, we used the SH spectra to determine shifts in the velocity of various lines. \citet{pereira10} studied the validity of the SH velocity fields by using synthetic spectra. They found that the distortion of the wavelength scale introduced by the telescope pointing uncertainties limits the accuracy of the velocity estimates up to $\pm10~\kms$, regardless of the S/N of the spectra, for sources which are neither point sources nor uniformly extended sources (e.g., the nuclear region of NGC~4945). Considering as well the uncertainty in the absolute wavelength calibration ($\sim10$\% of a pixel), which does depend on the S/N of individual spectra, \citeauthor{pereira10} estimated uncertainties of $10-30~\kms$ in the SH velocity fields, and concluded that variations of $>20~\kms$ in the velocity maps are likely to be real.

\begin{figure*}[htp]



  \hfill\includegraphics[width=14.2cm]{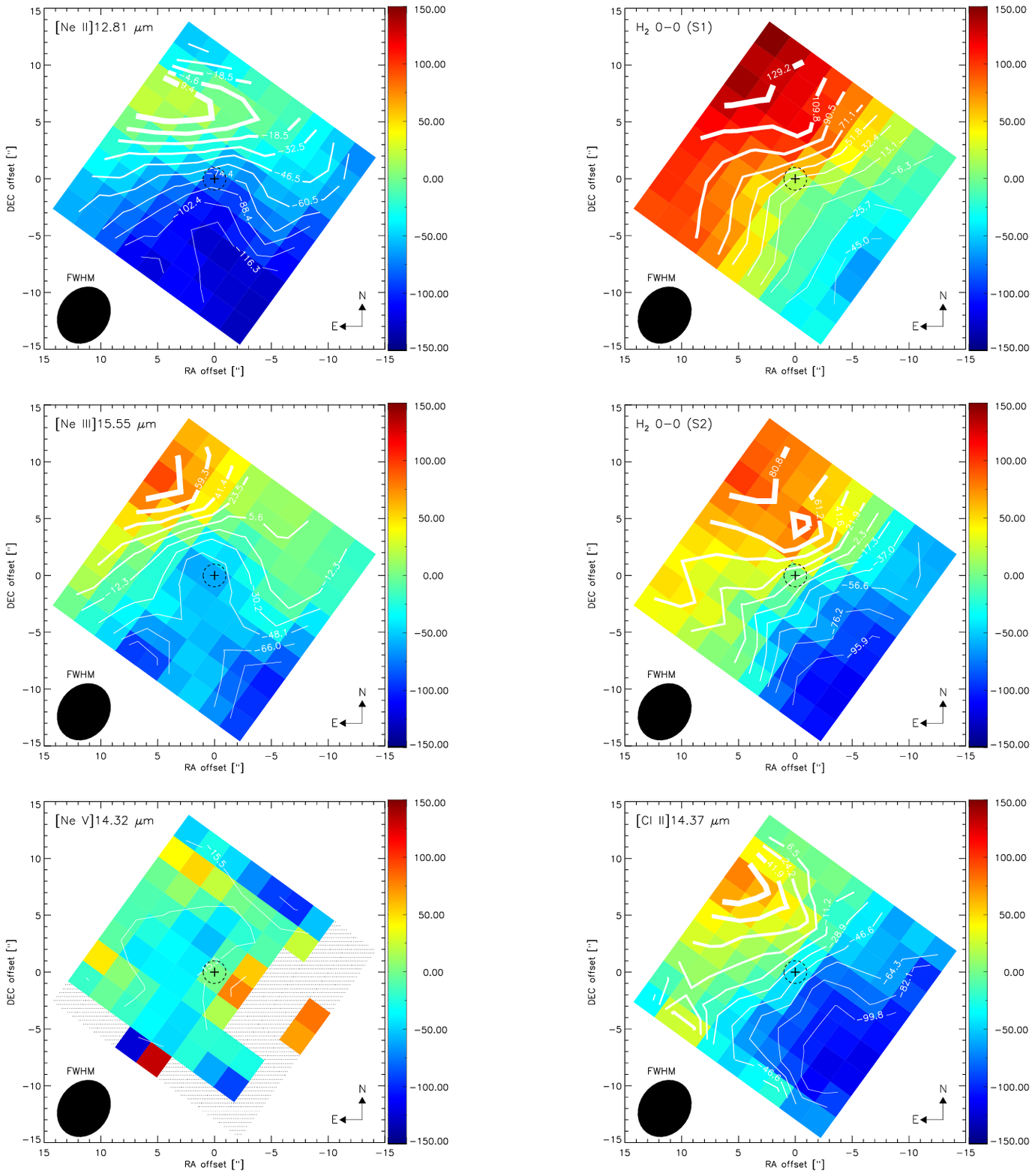}\hspace*{\fill}

  \caption{\footnotesize{The \textit{left panel} shows (from top to bottom) the velocity fields of the fine-structure lines \neii~12.81~\mum, \neiii~15.56~\mum, and \nev~14.32~\mum. The \textit{right panel} shows (from top to bottom) the molecular hydrogen lines H$_2$ 0-0 S(1) 17.0 \mum, and S(2) 12.3 \mum, and the \clii~14.37~\mum. The maps, covering a region of $\sim23''\times23''$ ($\sim426\times426~\rm pc^2$), are consistent with a rotating disk, with the exception of the \nev\ line, which present a relatively uniform (most likely not spectrally resolved) velocity field.}}
  \label{fig:rotation-maps}
\end{figure*}

The derived velocity fields of the \neii~12.81~\mum, \neiii\ 15.56~\mum, H$_2$ 0-0 S(1) 17.0 \mum\ and S(2) 12.3 \mum, and the \clii~14.37~\mum\ lines are shown in Fig.~\ref{fig:rotation-maps}. These correspond to the lines with the highest S/N ($\gtrsim100$) in the SH spectra throughout the whole region mapped, and therefore the velocity fields derived from them are considered to be reliable. We also include the velocity field of the \nev~14.32~\mum\ line, although with a much lower S/N ranging from $\sim1$ to $\sim100$ throughout the map. The pixels with less than a $3\sigma$ detection level in the \nev\ line (mostly found at the edges of the map, where the S/N is the lowest) are shown as hatched.

Given that the galaxy major axis runs at PA$\sim45^o$, and the fact that the inclination of the nuclear disk has been estimated to be smaller ($\sim62^o$) than the inclination ($\sim80^o$) of the large-scale galactic disk \citep{chou07}, we find that the range of velocities probed in a region of $\sim23''\times23''$ ($\sim426\times426~\rm pc^2$) with the \neii, \neiii, and H$_2$ S(2) lines agree with the range of velocities probed by the ISAAC long slit spectra of the \hi\ Pf$\beta$ and H$_2$ 0-0 S(9) lines by \citet[their Fig.7]{spoon03}.

The velocity structure of \neii\ (\textit{top left panel} in Fig.~\ref{fig:rotation-maps}) shows a gradient along the N-S direction rather than NE-SW, as in the case of H$_2$ 0-0 S(1) (\textit{top right panel} in Fig.~\ref{fig:rotation-maps}). A similar difference in velocity structure is observed in the \neiii\ and H$_2$ 0-0 S(2) velocity maps (\textit{middle left} and \textit{middle right} panels in Fig.~\ref{fig:rotation-maps}, respectively). 
These different kinematic components may be caused by different origins related to the species emitting, or by different levels of extinction (extinction law at different wavelengths), which allow us to probe deeper regions at one wavelength than another. This is further discussed in Sect.~\ref{sec:H2-gas}.

On the other hand, the relatively uniform central region of the \nev\ velocity field may be due to the lack of spectral resolution and the lower (factor 1 to 100) S/N level obtained in this line. This implies that we are unable to spectrally resolve, at a reliable level, the rather small ($\lesssim50~\kms$) velocity shifts in the \nev\ line. Nevertheless, with a higher spectral resolution we would expect a weaker rotation of \nev\ in comparison to the rotation shown by the starburst tracers (from about $-120~\kms$ to $\sim100~\kms$). If the \nev\ is exclusively related to the AGN NLR, this would mean that the NLR gas does not feel the gravitational pull of the large gas mass interior to the starburst ring. But it would feel the pull from the SMBH, showing not necessarily the same kinematics as the gas in the starburst ring. In fact, the group of three pixels next ($\sim43~\rm pc$) to the adopted location of the AGN (the H$_2$O mega maser), shows positive velocities (in the range $\sim10-70~\kms$ with uncertainties of $<$1\% in the line fit) which is opposite to the characteristics of the velocity fields of the starburst tracers. This might be a sign of the kinematically decoupled component discovered at the center of the disk with interferometric maps of the $J=2\rightarrow1$ transitions of \twco, \thco, and \ceo\ \citep{chou07}. However, higher spectral and spatial resolution observations of the \nev\ line are required to reliably conclude on this.

\section{Discussion}\label{sec:discuss}

\subsection{Tracing the circumnuclear starburst ring}\label{sec:starburst-ring}

The \neii\ and \siii\ fine-structure lines are our cleanest tracers of star formation (\hii\ regions). 
If an AGN is present, however, the \neiii\ and \siv\ lines will have contributions from both the starburst and the AGN. 
In NGC~4945 the AGN contamination to \neiii\ is likely low, given the faintness of \nev. In fact, at the peak flux of the \nev\ and \neiii\ maps (Fig.~\ref{fig:SH-maps1}) the former corresponds to only about 5\% of the \neiii\ flux.

From Table~\ref{tab-c4:fluxes-fov} we derive the respective luminosities of $L_{\rm [Ne V]}=7.42\times10^{29}~\Wsr$ and $L_{\rm [Ne III]}=1.76\times10^{31}~\Wsr$, using a luminosity distance $D_L$=3.82 Mpc. However, from the relation between $L_{\rm [Ne V]}$ and $L_{\rm [Ne III]}$ by \citet[][]{gorjian07} (which holds for a group of Seyfert 1 and 2, AGNs, ULIRGs and radio galaxies), the predicted \neiii\ luminosity should be $\sim3.28\times10^{30}~\Wsr$, that is a factor $\sim$5.4 lower than observed.
This means that NGC~4945 has a slightly lower AGN contribution to the \neiii\ line than the outlier NGC~3079 found in the sample by \citet[][]{gorjian07}, for which the \neiii\ luminosity is a factor $\sim$5 brighter than expected based on its \nev\ luminosity.



\begin{figure*}[!htp]
  




  \hfill\includegraphics[width=14.2cm]{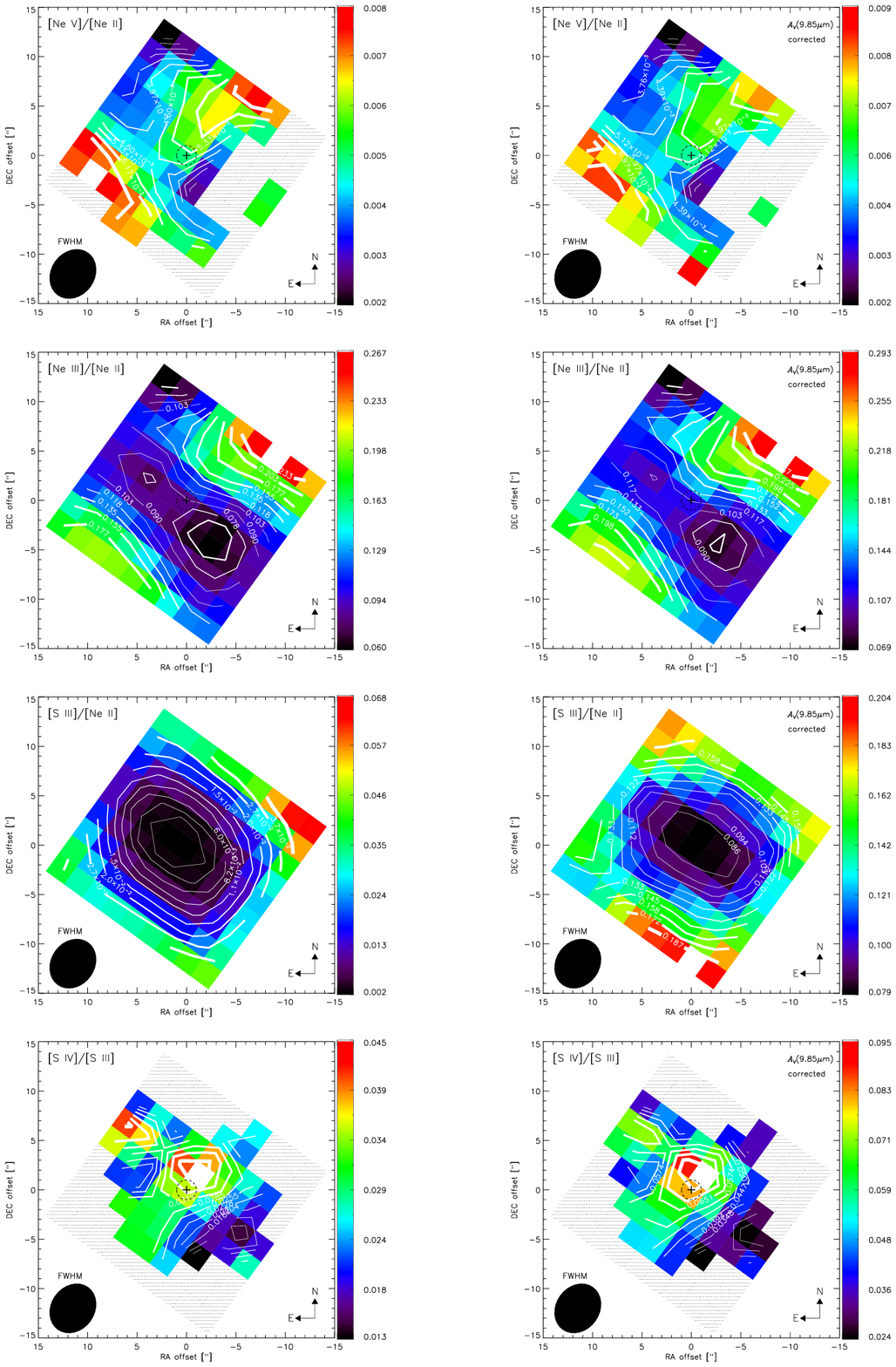}\hspace*{\fill}

  \caption{\footnotesize{\textit{Left panels} - Integrated flux ratios between the fine-structure lines (from top to bottom) \nev~14.32/\neii~12.81, \neiii~15.56/\neii~12.81, \siii~18.71/\neii~12.81, and \siv~10.51/\siii~18.71.
\textit{Right panels} - Same as above, but considering the fluxes corrected for extinction. 
The hatched areas correspond to pixels with a $<$3$\sigma$ detection level in the faintest emission lines \nev~14.32~\mum\ and \siv~10.51~\mum.}}
  \label{fig:line-ratios}
\end{figure*}

The \neiii~15.56/\neii~12.81 \mum\ line ratio (Fig.~\ref{fig:line-ratios}) ranges between $\sim$0.13 and $\sim$0.27 above and below the major axis of the region mapped. The difference with the lowest ratios (0.06--0.13), which are found along the northeast-southwest axis, is more pronounced than in the \nev~14.32/\neii~12.81 \mum\ line ratios (\textit{top left panel} in Fig.~\ref{fig:line-ratios}). 

\begin{figure}[tp]



  



  \hfill\includegraphics[width=6.5cm]{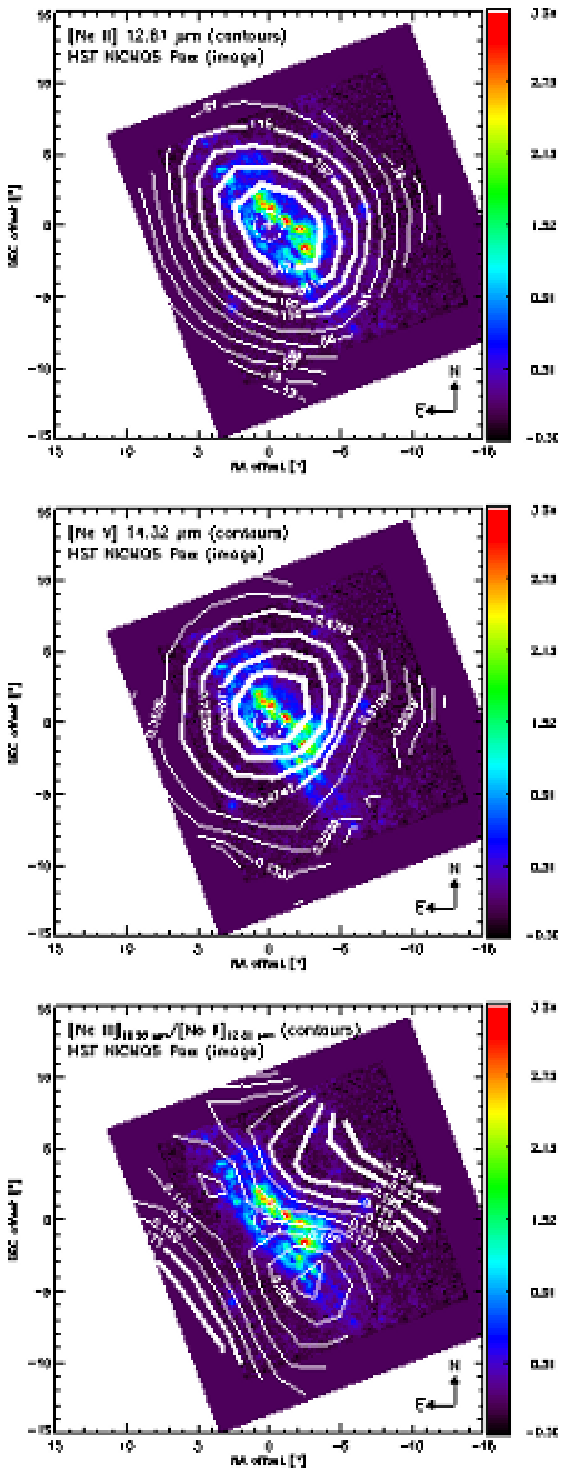}\hspace*{\fill}

  \caption{\footnotesize{HST NICMOS image of Pa$\alpha$ (in units of $10^{-21}~\Wcm$) of the nucleus of NGC~4945 \citep{marconi00}. The not corrected for extinction contours (labelled in units of $10^{-12}~\Wcmsr$) are the fine-structure lines \neii~12.81~\mum\ (\textit{top panel}) and \nev~14.32~\mum\ (\textit{middle panel}). 
The \nev\ flux peaks at about one pixel to the northwest of the H$_2$O mega maser (Fig.~\ref{fig:SH-maps1}), 
The \textit{bottom panel} shows the \neiii~15.55/\neii~12.81~\mum\ line ratio not corrected for extinction. The ratio is lower along the starburst ring traced by the Pa$\alpha$ line, while it increases away from the center.}}
  \label{fig:PAalpha}
\end{figure}

The lowest \neiii~15.55~\mum/\neii~12.81~\mum\ line ratios coincide with the 100 pc-scale circumnuclear starburst ring of NGC~4945 (\textit{bottom panels} in Fig.~\ref{fig:PAalpha}), detected in Pa$\alpha$ emission with HST NICMOS \citep{marconi00}. 
The increasing gradient above the starburst ring seems to mimic the conical-shaped cavity traced by the H$_2$ 1-0 S(1) line \citep{marconi00}. However, the ratio also increases toward the south-east direction from the starburst ring, which rather implies that the ring has different properties than its surroundings. 

In fact, a similar increase of \neiii/\neii\ ratios away from the nucleus is observed in the nuclear maps by \citet[][their Fig.5]{pereira10}, with exception of NGC~7130, the only galaxy classified as LINER/Seyfert 2 in their sample. 
According to \citep{snijders07} the higher the metallicity, the higher the age of the stellar population and the higher the gas density, the lower should be the \neiii~15.55~\mum/\neii~12.81~\mum\ ratio.

The \siii~18.71/\neii~12.81 line ratio is considered a good tracer of densities in the range $10^4$ and $10^6~\3cm$ because their similar excitation potentials (21.6 eV and 23.3 eV, for \neii\ and \siii, respectively) and significantly different critical densities ($6.1\times10^5~\3cm$ for \neii\ and $1.0\times10^4~\3cm$ for \siii) make this ratio less sensitive to the hardness of the radiation field than to the density of the ISM. According to the models by \citet{snijders07} a lower \siii/\neii\ ratio would indicate larger densities. 
In Fig.~\ref{fig:line-ratios} the \siii~18.71/\neii~12.81 line ratios are also lower in the center than around it. This ratio ranges between 0.002 and 0.013 in the center, which indicate densities larger than $10^6~\3cm$ in a $>$5 Myr old starburst system with solar metallicity and relatively high ($\rm q=8\times10^8$) ionization parameter \citep[][their Fig.5]{snijders07}.
Based on our observed \neiii/\neii\ ratios, the \siii/\neii\ ratios predicted with the model by \citet[][their equation 2]{pereira10} are about ten times larger than the observed ratios. These can be a consequence of the about 10 times higher extinction found in NGC~4945 than in the sample of galaxies used by \citet{pereira10}.

The \neiii/\neii\ line ratios obtained with the extinction correction (\textit{middle right panel} in Fig.~\ref{fig:line-ratios}) are just $\sim$9\% larger than without correction. This relatively small change after the extinction correction is because even in a high-extinction situation the differential extinction between \neiii\ and \neii\ is small, given that both lines are closely spaced in wavelength and not in one of the silicate absorption features. On the other hand, because of their larger differential (wavelength) extinction, the \siii/\neii\ ratios corrected for extinction do change significantly from a factor $\sim$50 in the center (where the extinction is larger) to a factor 3 away from the center (where the extinction is lower).



Note that only the ratios from our co-added fluxes (Table~\ref{tab-c4:fluxes-fov}) can be compared to other galactic nuclei, as the 10$\times$10 aperture is comparable to the size probed in any of the more distant galaxy nuclei. 
The \neiii/\neii\ ratio at the position of the H$_2$O mega maser is about 10\% lower than the ratio obtained from the fluxes of the 10$\times$10 co-added spectrum (Table~\ref{tab-c4:fluxes-fov}).

Since in starburst environments the \neii\ and \neiii\ emission lines are expected to be driven mainly by photo-ionization \citep[e.g.,][]{ho07}, the lower ratios found along the northeast-southwest axis are likely due to a \neii\ emission enhanced by the starburst ring. Although, these low ratios are also consistent with a ratio \neiii/\neii$\leq$0.1 found in shocks \citep{binette85} where the low ionization line \neii\ can also be enhanced \citep{voit92}. 

On the other hand, the highest \neiii/\neii\ ratios found above and below the starburst ring are larger than those typically found in shocks. These higher ratios could 
actually be tracing an additional contribution to the 
[NeIII] emission by a conically shaped narrow line 
region (NLR). Previously, no evidence was found for the
existence of such a NLR, given the absence of 5007\AA\ \oiii\ 
emission in the central 800pc$\times$800pc \citep{moorwood96a}. 
Instead the conical cavity traced by optical and
near-infrared line and continuum tracers \citep{moorwood96a, marconi00} was associated with a starburst 
super wind. However, given the high extinction implied
by our observations, optical \oiii\ emission can be 
easily attenuated; mid-infrared \neiii\ emission less so. 
A hypothetical conical NLR would be difficult to identify 
from a \neiii\ map alone due to blending with the nuclear 
starburst component. But it may manifest itself more 
clearly in a \neiii/\neii\ map, in which the contribution 
of the starburst component diminishes quickly beyond the 
nucleus proper.

The \neiii/\neii$\sim$0.06--0.13 ratios observed along the starburst ring are in close agreement with the ratios observed in some galaxies of the ISO starburst sample and are consistent with burst timescale models predicting a relatively short-lived starburst of $5-8~\rm Myr$ \citep[their Fig.6]{thornley00}. 
In the whole region mapped, we  observe ratios \neiii/\neii$<$0.3, which are a factor $\sim$3 lower than the ratios observed in the sample of quasars reported by \citet{veilleux09}. However, the ratio \neiii/\neii$\sim$0.07 observed at the nucleus of NGC~4945 is comparable to the ratios observed in the nucleus of 5 (out of 16) galaxies in the sample of LIRGs studied in \citet[their Fig.14]{pereira10}. 

On the other hand, the differential extinction between \siii\ and \siv\ is larger than between \neii\ and \neiii.
Because \siv\ is sitting close to the deepest point of the silicate absorption feature at 9.7~\mum\ (Fig.\ref{fig:average-FOV-spectra}), the \siv~10.51/\siii~18.71 \mum\ line ratios corrected for extinction (\textit{bottom right} in Fig.~\ref{fig:line-ratios}) are a factor $\sim$2 larger than those obtained without correction (\textit{bottom left} in Fig.~\ref{fig:line-ratios}). In both cases the highest ratio is found about $2.3''$ (one pixel) northwest of the H$_2$O mega maser.
The uncorrected for extinction map-averaged ratios (from the fluxes in Table~\ref{tab-c4:fluxes-fov}) of \neiii/\neii$\sim$0.11 and \siv/\siii$\sim$0.02 are similar (within 10\%) to the ratios obtained from the SH staring observations by \citet{bernards09}. They position NGC~4945 among the sources with the lowest hardness of the radiation field in their sample of starburst galaxies, which, according to \citeauthor{bernards09}, may be an indication of an old (or small) population of massive stars in NGC~4945. However, this can be concluded only for the outermost surface that we can probe with \siv\ and \siii\ lines, and we cannot rely much on the extinction corrected data for this analysis, since the \siv/\siii\ ratio depends more on the extinction law used than the \neiii/\neii\ ratio.

\subsection{Molecular hydrogen gas}\label{sec:H2-gas}

\begin{figure*}[!tp]


  \hfill\includegraphics[width=14.2cm]{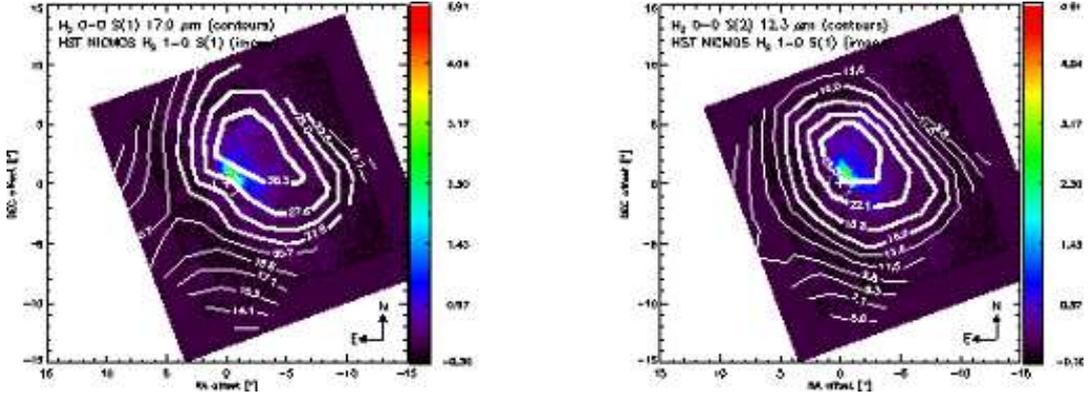}\hspace*{\fill}

  \caption{\footnotesize{IRS/SH contours of the rotational H$_2$ 0-0 S(1) (\textit{left}) and H$_2$ 0-0 S(2) (\textit{right}) lines overlaid on the HST NICMOS map of the vibrational H$_2$ 1-0 S(1) molecular hydrogen line \citep{marconi00}.}}
  \label{fig:HST-H2-IRS-H2}
\end{figure*}

As can be seen in Figures \ref{fig:SH-maps1} \& \ref{fig:SH-maps2}, the emission centroids of the H$_2$ 0-0 S(1), S(2) and S(3) lines are clearly offset from the centroids of the low ionization fine-structure lines, which all peak close to the H$_2$O mega maser. Such an offset is hard to understand if the H$_2$ 0-0 S(1), S(2) and S(3) lines and the HII region gas would be exposed to similar amounts of extinction. The differential extinction between H$_2$ S(2) and \neii, at 12.28 and 12.81 \mum, for instance, is very small. The same is true for the wavelengths of H$_2$ S(1) and \siii\ at 17.0 and 18.71 \mum, respectively. The lack of H$_2$ 0-0 S(1), S(2) and S(3) emission at the position of the mega maser can be explained if the molecular hydrogen emission in the nucleus would be more strongly attenuated than the HII region gas. That is, the inner H$_2$ emission would suffer additional/differential extinction that is not accounted for by the silicate feature. 
On the other hand, analysis of the higher 
pure rotational lines observed in the SL module
(Spoon \etal\ \textit{in preparation}) and with VLT-ISAAC
\citep[][their Figure 7]{spoon03} indicates that 
the S(5), S(7) and S(9) lines originate progressively 
closer to the nucleus. This seems hard to reconcile
with the above scenario, given that A($\lambda$)/A$_V$
in the S(5)-S(9) wavelength range is likely higher 
than in the S(1)-S(2) wavelength range.

We instead favour a scenario in which the lower
pure rotational lines S(1) and S(2) mainly originate
from an unobscured extra-nuclear component associated 
with the super-wind cone as seen in the HST NICMOS map 
of the vibrational H$_2$ 1-0 S(1) emission \citep{marconi00}. 
Figure~\ref{fig:HST-H2-IRS-H2} shows the IRS/SH maps of the pure
rotational H$_2$ S(1) and S(2) lines overlaid on the H$_2$
1-0 S(1) map. Interestingly, the two pure rotational 
lines peak further from the apex of the cone than 
the 1-0 S(1) line does. The 0-0 S(1) line (upper 
level energy 1015 K) more so than the 0-0 S(2) line 
($T_{up}$=1682 K). 
In this scenario the higher pure rotational lines
S(5), S(7) and S(9) would originate mainly from the
starburst ring or within.
There one likely finds the high critical densities ($n_{\rm crit}>10^5~\3cm$) 
as well as high gas temperatures ($T_{\rm K}>10^3$ K) needed to thermalize these lines 
\citep[as seen in active galaxies like NGC~1068, Mrk~231, and Arp~220; e.g. ][]{pb07, pb09, aalto07a, vdwerf10}, 
but they will also be affected by strong extinction. 
The latter may cause the nuclear 
component of the 2.12 $\mu$m H$_2$ 1-0 S(1) line 
($T_{up}$=6950 K) to suffer far more extinction 
than the longer wavelength H$_2$ 0-0 S(5)-S(9) lines 
and therefore to be associated solely with the conical 
cavity. In all of this the H$_2$ 0-0 S(3) line, originating 
at an upper level energy of 2500 K, may have both strong 
nuclear and cavity components. Due to the extreme
extinction the nuclear component would be suppressed,
leaving only the cavity component for us to see.

\subsection{Excitation of H$_2$}\label{sec:H2-Tex}

For LTE conditions, and an ortho-to-para ratio of 3 \citep[their Fig.13]{neufeld06}, we can estimate the excitation temperature of the molecular hydrogen throughout the region mapped with the IRS/SH module, from the ratio between the flux of the H$_2$ 0-0 S(2) 12.3 \mum\ and H$_2$ 0-0 S(1) 17.0 \mum\ lines as 

\begin{equation}\label{eq:H2-Tex}
 T_{ex}=-\frac{T_{up}^{\rm S(2)} - T_{up}^{\rm S(1)}}{ log(F_{\rm S(2)}\nu_{\rm S(1)}A_{\rm S(1)}g_{\rm S(1)}) - log(F_{\rm S(1)}\nu_{\rm S(2)}A_{\rm S(2)}g_{\rm S(2)}) }~~~\rm K,
\end{equation}

\begin{figure*}[!tp]


  \hfill\includegraphics[width=14.2cm]{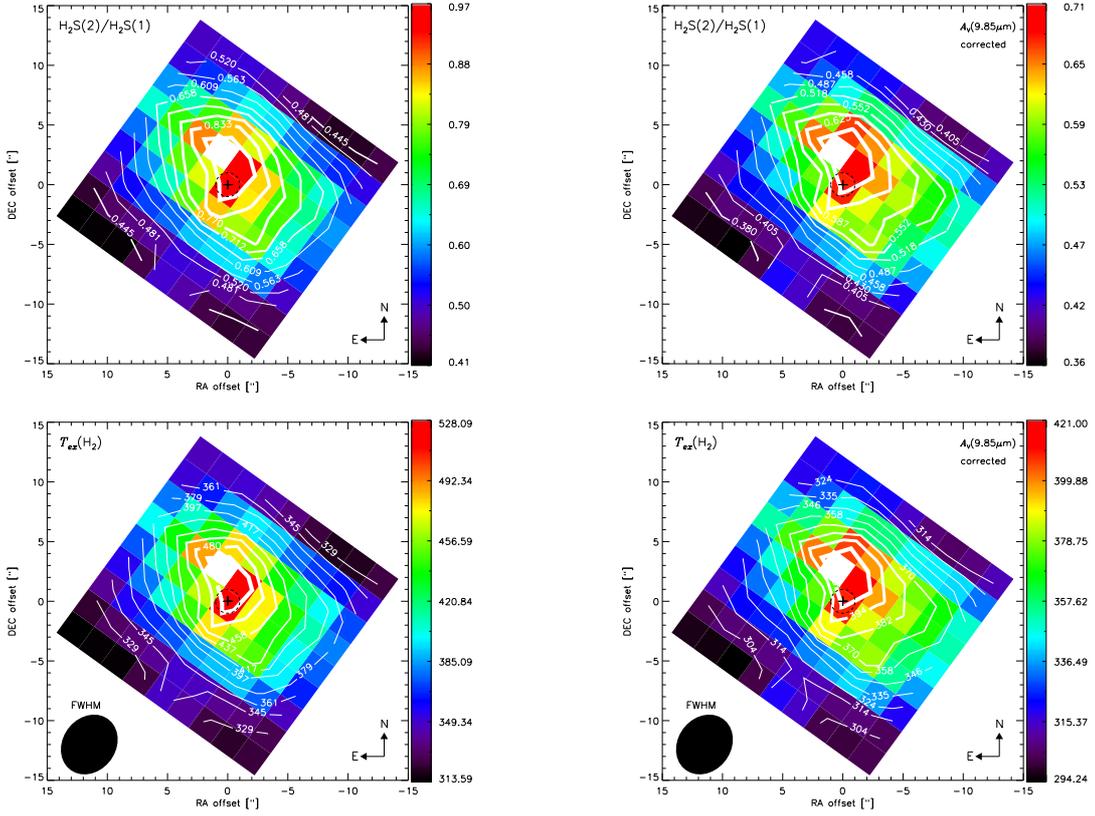}\hspace*{\fill}

  \caption{\footnotesize{\textit{Left panels} - IRS/SH maps of the H$_2$ 0-0 S(2) to H$_2$ 0-0 S(1) total flux ratio (\textit{top left panel}), and the map of the estimated $T_{ex}$ ($\rm K$) of the molecular hydrogen. The peak excitation temperature of $\sim$528$~\rm K$ is reached about $\gtrsim$2.3$''$ (one pixel) to the north of the H$_2$O mega maser. \textit{Right panels} - Same as above, but using the extinction corrected fluxes of the H$_2$ S(2) and S(1) lines. The extinction correction leads to an $\sim$100$~\rm K$ lower peak temperature.}}
  \label{fig:H2-Tex}
\end{figure*}

\noindent
where $F_{\rm S(1)}$ and $F_{\rm S(2)}$ are the integrated flux densities of the H$_2$ 0-0 S(1) and S(2) lines, $\nu$ is the corresponding rest frequency ($\sim1.75\times10^{13}~\rm Hz$ for S(1) and $\sim2.44\times10^{13}~\rm Hz$ for S(2)), and $A$ and $g$ are the respective Einstein $A$-coefficients and statistical weights of each transition. The upper level energy $T_{up}$ of the transitions is in units of $\rm K$.

Figure~\ref{fig:H2-Tex} shows the IRS/SH maps (not corrected for extinction) of the H$_2$ 0-0 S(2) to S(1) total flux ratio (\textit{top left panel}), and the corresponding map of the excitation temperature $T_{ex}$ ($\rm K$) (\textit{bottom left panel}) of the molecular hydrogen, estimated from eq.(\ref{eq:H2-Tex}). The peak excitation temperature of $\sim528\pm31~\rm K$ is reached $\gtrsim$2.3$''$ (one pixel) to the north of the H$_2$O mega maser. The \textit{right panels} of Fig.~\ref{fig:H2-Tex} show the H$_2$ 0-0 S(2)/S(1) ratio and the estimated excitation temperature from the respective H$_2$ fluxes corrected for extinction with the silicate-based $A_{\rm V}(9.85~\mu \rm m)$. The correction for extinction leads to a peak temperature of $421~\rm K$, which is about $100~\rm K$ lower than the temperature derived from the non-corrected fluxes. However, the distribution of the temperature, and the position of its peak value, do not change.

Since galaxies are complex systems, the H$_2$ gas is not expected to be at a single temperature. Besides, a long line of sight can probe different excitation environments (e.g., PDRs, XDRs, shocks) like in the sample of galaxies studied by \citet{roussel07}.
Previous estimates of the excitation temperature, based on SWS observations of the (0-0) S(0) and S(1) fluxes detected in the nucleus of NGC~4945, led to a cooler H$_2$ component with $T_{ex}\sim160~\rm K$ (corresponding to about 9\% of the total H$_2$ mass), while a temperature $T_{ex}\sim380~\rm K$ (about 0.4\% of H$_2$ mass) was estimated from the S(1) and S(2) fluxes \citep[their Table 6]{spoon00}, which is in close agreement with our result considering the size of the big SWS slit. This excitation temperature is similar to the temperature $T_{ex} = 365\pm50~\rm K$ derived for the low-energy transitions in NGC~1377 \citep{roussel06}, but higher than the temperature of $T_{ex} = 292\pm6~\rm K$ estimated from the S(0)-S(3) lines observed in NGC~6240 \citep{higdon06}.

Since we cannot use the map of the H$_2$ S(3) 9.7 \mum\ line due to the flux mismatch between the SH and SL modules mentioned in Sec.~\ref{sec:results}, we would only be able to base our estimate of the warm H2 mass on the S(1) and S(2) lines. In fact, even if we had been able to use the S(3) line, most of the mass is revealed by including the S(0) line. This line has however not been mapped, since the spatial resolution of the LH module is much lower than that of the SH and SL modules.

\subsection{AGN dominated [Ne V] emission?}\label{sec:NeV-dominated}

The detection of \nev\ emission from the nucleus of NGC4945
is not surprising. The galaxy hosts a 1.4$\times$10$^6$ M$_{\odot}$ 
super massive black hole, as revealed by H$_2$O maser observations 
\citep{greenhill97} with an intrinsic 0.1--200\,keV luminosity of 
1.8$\times$10$^{43}$ erg s$^{-1}$ \citep{guainazzi00}.

More remarkable is the faintness of the \nev\ emission. 
Only 10$^{-6}$ of the infrared luminosity of the galaxy is 
detected in the \nev\ line (Table~\ref{tab-c4:fluxes-fov}). This is surprisingly 
little for an AGN which can easily account for the entire 
bolometric emission of the galaxy \citep{marconi00}. It is also
100 times less than L(\nev)/L$_{\rm IR}$ for galaxies whose 
IR luminosity is dominated by AGN activity \citep{goulding09}.

Interesting is also the size of the \nev\ emitting region,
which can be reasonably well constrained given that at a
distance of 3.82\,Mpc one arcsecond corresponds to 18.5\,pc.
Since the spatial profile of the \nev\ map (Figure~\ref{fig:SH-maps1}) is only
marginally wider than the PSF, the radius of the \nev\ emitting 
region will be less than 55 pc ($\sim$3$''$).
This upper limit is large in comparison to the
size of coronal line regions \citep[e.g. $\sim$19 pc for 
Circinus as probed by the 2.48\mum\ \SiIV\ line;][]{prieto04}, but small compared to the kpc scale of 
narrow line regions as probed with \oiii\ \citep[e.g. Circinus;][]{marconi94}.

Both the faintness and the observed extent of the \nev\ emission
are likely affected by strong extinction within the nucleus.
However, if the narrow line region were to extend above 
the plane of the molecular disk in which the AGN and the
circumnuclear starburst are embedded, the amount of extinction
would be less than probed by the silicate absorption feature 
($A_{\rm V}$=50-60). 
The amount will, however, certainly be less than the line 
of sight extinction to the AGN broad line region,
N$_H$=2.4$\times$10$^{24}$ cm$^{-2}$, \citep{guainazzi00}, 
which is equivalent to $A_{\rm V}$=1300.

The extinction probed by the silicate absorption feature is consistent with the F(\nev)/F(14-195\,keV) ratio found for Sy1 galaxies 
\citep{weaver10} when multiplied by the absorption corrected 14-195\,keV flux for NGC\,4945.
The latter depends on the geometry of the absorber, and ranges between 5 and 10 times the Swift-BAT flux of NGC\,4945 (F(14-195\,keV)=33$\pm$0.8 $\times$10$^{-18}$ W cm$^{-2}$; \citet{tueller10}) for a spherical and edge-on disk geometry, respectively (K. Iwasawa, private communication). 
Taking further into account the factor 2.5 scatter in the F(\nev)/F(14-195\,keV) ratio for Sy1 galaxies, the attenuation on the \nev\ flux ranges between 12 and 157, which corresponds to $A_{\rm V}$=55-112 mag for our adopted local ISM extinction law of \citet{chiar06}. 
This result would indicate that the NLR is buried along with the other components of the nuclear molecular disk.

The strong obscuration to the NLR in NGC\,4945 may serve
as a warning that mid-infrared NLR tracers such
as \nev\ and \oiv\ may not always be suitable as tracers
of the AGN luminosity as suggested by \citet{goulding09}. 
Galaxies in this risk group may be recognized by the
presence of a deep silicate absorption feature in their
mid-infrared spectra and a classification 3A, 3B or 3C
in the mid-infrared galaxy classification scheme of \citet{spoon07}.
Besides NGC\,4945 galaxies in this group include LIRGs such
as NGC\,4418 and ULIRGs like Arp220, IRAS\,08572+3915 and 
IRAS\,F00183-7111.

The example of NGC~4945 may also serve as a warning
that the non-detection of NLR tracers in AGN does not 
necessarily imply the absence of a NLR. Instead it may 
indicate that the NLR is so strongly attenuated that
the line does not emerge above the continuum noise. 
In these cases, only deep mid-IR observations, like 
the ones presented here, would reveal these tracers.

In NGC4945 the \nev\ line was detected at 5\% of the 
continuum flux in a spectrum with continuum S/N of 
140 in the 14-15\mum\ range. Other galaxies which 
may host an AGN, and which have been observed at 
similar S/N (among them ULIRGs like Arp220 and 
IRAS\,08572+3915), do not show evidence for \nev\ 
emission \citep{armus07}. The NLR in these galaxies may 
hence be even more strongly obscured than in NGC~4945, 
or the AGN may contribute a far smaller fraction of the 
bolometric power than in NGC~4945. The NLR may also 
simply not exist if the ionizing photons needed to 
form the NLR are 4$\pi$ obscured close to their 
origin.

\subsection{[Ne V] emission from supernovae? the template Cassiopeia A}\label{sec:NeV-SNR}

Although the [NeV] emission is unresolved along the major axis and only marginally resolved along the minor axis (as discussed in Sec.~\ref{sec:results}), the fact that the \nev\ line has also been observed in supernova remnants (SNRs) \citep[e.g.][]{oliva99, smith09} raises valid questions: could the \nev\ emission observed in NGC~4945 be powered purely by SNRs?, and if so, how many SNRs would be needed to reproduce its \nev\ flux?
Below we explore the alternative of \nev\ emission being powered by the SNRs in the nuclear region of NGC~4945. We note, though, that the number of supernovae needed depends not only on the total \nev\ flux but also on the actual size of the SNR.

Our analysis is based on previous Spitzer/IRS SL module observations of Cassiopeia A (Cas A) from which the global distribution of fine-structure lines covering a 5.3$\times$5.3 arcminute$^2$ area was presented by \citet{smith09}.
The total \nev\ flux of Cas A was not reported before due to the difficulties to extract the flux with enough reliability.
Using fitting techniques tailored to the \nev\ line, we now are able to state its total flux as $1.5\pm0.7\times10^{-18}~\Wcm$. 
Considering an average extinction $A_{\rm V}\sim5$ mag \citep{hurford96} the corrected flux, using the extinction law for the local ISM by \citet{chiar06}, is $\sim1.9\pm0.9\times10^{-18}~\Wcm$. 

On the other hand, the total \nev\ flux in NGC~4945 is $\sim5.3\pm0.5\times10^{-21}~\Wcm$. In order to estimate the dereddened \nev\ emission observed in NGC~4945, we computed an average extinction value $A_{\rm V}\sim36$ mag from the $A_{\rm V}(9.85~\mu \rm m)$ map (Fig.~\ref{fig:Tau-sil}). Note that due to the stronger extinction at the central pixels, this represents just a lower limit to the actual extinction on the \nev\ line. 
Thus the lower limit for the extinction-corrected total \nev\ flux is $\sim$27.9$\times10^{-21}~\Wcm$.

Considering a canonical distance of 3.4 kpc for Cas A \citep[][and references therein]{smith09} and $\sim$3.82 Mpc for NGC~4945, we found that the ratio between the total \nev\ flux of NGC~4945 and Cas A, corrected by the square of the ratio between the distances $(D_{\rm NGC4945}/D_{\rm CasA})^2$, is about 19$\times10^3$. This represents a lower limit to the number of contemporaneous Cas A type supernovae needed to produce the extinction-corrected \nev\ emission observed in NGC~4945. Another extreme lower limit can be obtained assuming no extinction in the NGC~4945 flux, from which the same exercise indicates that about 4$\times10^3$ Cas A type SNRs would be needed.

According to \citet{lenc09}, the upper limit to the supernova rate in NGC~4945 is 15.3 $\rm yr^{-1}$, which means the individual supernovae, for the not corrected and corrected for extinction lower limits estimated above, need to emit at Cas A levels for about 260 and 1240 years, respectively.
These are reasonable numbers given that the age of Cas A is estimated to be about 330 yr \citep{fesen06}, while 
the kinematic age of another \nev\ emitting SNR, E0102, has been estimated to be $2050\pm600$ yr \citep{finkelstein06}. This implies that the number of SNRs estimated above can indeed persist long enough to power the \nev\ emission observed in NGC~4945, at least from the point of view of the SNR properties used in our estimate. However, \citet[][their section 4.5]{lenc09} also determine a median SN age of 85 yr, which translates into a median supernova rate of $\sim$0.12 SNe/yr, or a much (15.3/0.12$\sim$128 times) longer radiative phase ($\sim$3--16$\times10^4$ years) needed for the individual supernovae. 
Therefore, the \nev\ flux observed in NGC~4945 is likely not powered by a population of supernova remnants.

\section{Final remarks}\label{sec:final-remarks}

We have mapped the central region of NGC~4945 with the SH and SL modules of the Spitzer InfraRed Spectrograph.
From the SH spectral cubes we produced maps of fine-structure emission lines \siv\ at 10.51 \mum, \neii\ at 12.81 \mum, \clii\ at 14.37~\mum, \neiii\ at 15.56 \mum, \siii\ at 18.71 \mum, the AGN narrow-line region tracer \nev\ at 14.32 \mum, and the molecular hydrogen lines, H$_2$ S(2) and H$_2$ S(1) at 12.3 \mum\ and 17.0 \mum, respectively. 
From the SL spectral cubes we obtained maps of the H$_2$ S(3), and the silicate absorption feature at 9.7~\mum.

We present the first map of \nev~14.32~\mum\ towards the nucleus of NGC~4945 with flux detection levels down to $0.1\times10^{-12}~\Wcmsr$ per pixel. We produced and estimated an extinction map $A_{\rm V}(9.85~\mu \rm m)$ based on the apparent strength of the 9.7~\mum\ silicate absorption feature, which traces the contours of the starburst ring at a $\sim$5$''$ spatial resolution. 

Most of the emission lines are found to peak on the nucleus, within the uncertainty of the astrometry. Only the warm molecular hydrogen emission shows a maximum about 60--100 pc NW of the nucleus. 
After correction for extinction the distribution of the H$_2$ rotational emission is more concentrated in the nuclear region, but its peak emission is still slightly offset from the peak of the other emission lines, that of the H$_2$O mega maser. 
Thus, we favour a scenario in which the lower pure rotational lines S(1) and S(2) mainly originate from an unobscured extra-nuclear component associated 
with the super-wind cone as seen in the HST NICMOS map of the vibrational H$_2$ 1-0 S(1) emission, with an intrinsic excitation trend toward the nucleus which is reflected in the higher level S(5)-S(9) lines.

We found that the map-integrated \nev/\neii\ ratio is consistent with ratios observed in starbursts rather than in AGNs. 
The \neiii/\neii$<$0.13 ratios observed along the starburst ring are likely due to an excess \neii\ emission driven by the starburst ring, or to high density ($>10^6~\3cm$) ISM gas in the circumnuclear disk.

A range of extinction $A_{\rm V}\sim55-112~\rm mag$ (which corresponds to an attenuation of a factor 12--160) estimated for \nev\ from our observed \nev\ flux and the absorption-corrected 14-195\,keV Swift-BAT flux, indicate that mid-infrared NLR tracers such as \nev\ and \oiv\ may not be trusted as tracers of the AGN luminosity of galaxies with a deep silicate absorption feature.

A new estimate of the total \nev\ flux in Cassiopeia A indicates that at least 4--19$\times10^3$ Cas-A type supernova remnants, with ages between $>$260 (no extinction correction) and $>$1240 (on extinction correction) years, would be needed to power the \nev\ emission observed in NGC~4945. However, given the actual median age of SNRs observed in NGC~4945, and the uncertainty in the true extinction of its \nev\ emission, SNRs are not likely to fully reproduce the \nev\ flux observed in the nucleus of this galaxy.

\begin{acknowledgements}
We thank Varoujian Gorjian for the Spitzer-MIPS 24 \mum\ map of the central region of NGC~4945.
We also thank Alessandro Marconi for providing the HST NICMOS images, and Emil Lenc for the ATCA 2.3 GHz image. 
We thank Kazushi Iwasawa for the very useful comments and the correction factors provided for the SWIFT-BAT fluxes.
We are grateful to Aleks Diamond-Stanic for constructive discussions. 
We also thank the referee for his/her pertinent and insightful comments.
We are also grateful to the SPITZER/SINGS team for their support during and after the observations. 
\end{acknowledgements}

\bibliographystyle{aa}
\setlength{\bibsep}{-2.1pt}
\bibliography{jp}





\end{document}